\documentclass[preprint2]{aastex}
\shorttitle{COMPARATIVE STUDY OF ASYMMETRY ORIGIN OF GALAXIES IN DIFFERENT ENVIRONMENTS}
\shortauthors{Plauchu-Frayn \& Coziol}

\begin{document}

\title{Comparative Study of Asymmetry Origin of Galaxies in \\ Different Environments.
I. Optical observations.}

\author{I. Plauchu-Frayn \altaffilmark{1} \& R. Coziol \altaffilmark{1}}
\email{plauchuf@astro.ugto.mx, rcoziol@astro.ugto.mx}
\altaffiltext{1}{Departamento de Astronom\'{\i}a, Universidad de Guanajuato Apartado Postal 144, 36000 \\Guanajuato, Gto, M\'exico}

\begin{abstract}

This paper presents the first of two analyses about the influence of
environment on the formation and evolution of galaxies observed in
the nearby universe. For our study, we used three different
samples representing different density environments: galaxies in
Compact Groups (HCGs), Isolated Pairs of Galaxies (KPGs),  and Isolated
Galaxies (KIGs), which were taken as references. Using both
characteristic isophotal parameters and evidence of asymmetries in the
optical and the near-infrared, we are able to establish differences in
the characteristics of galaxies with different morphologies in
different environments, allowing us to better understand their
different formation histories. In this first paper we present the isophotal and asymmetry
analyses of a sample of 214 galaxies in different environments
observed in the optical ($V$ and $I$ images). For each galaxy, we
have determined different characteristic isophotal parameters and $V
- I$ color profiles, as a function of semi-major axis, and performed
a full asymmetry analysis in residual images using the $V$ filter.
Evidence of asymmetry in the optical is almost missing in the KIG
sample, and significantly more common in the KPG than in the HCG
samples. Our isophotal analysis suggests that the stellar
populations in the HCG galaxies are older and more dynamically
relaxed than in the KPG. The HCG galaxies seem to be at a
more advanced stage of interaction than the KPGs. One
possible explanation is that these structures formed at different
epochs: compact groups of galaxies would have formed before close
pairs of galaxies, which only began interacting recently. However,
similarities in the formation process of galaxies with same
morphology suggest CGs and close pairs of galaxies share similar
conditions; they are new structures forming relatively late in low-density environments.

\end{abstract}


\section{Introduction}

Although it is recognized that the environment of galaxies plays an
important role in their formation and evolution, the mechanisms
responsible for such processes, the details on how they apply, and
the time-scales on which they are effective are still largely
unknown. For example, in Compact Groups of Galaxies (CGs) we have
recently shown \citep{cp07} that mergers and tidal
interactions are two important mechanisms driving the morphological
evolution of galaxies in these systems. We have also found that many
of the ongoing merger events were possibly happening without gas, a
phenomenon known in the literature as a ``dry merger'' \citep{vandokkum05}. 
According to the dry merger hypothesis, elliptical galaxies are
generally formed by the merger of bulge-dominated galaxies, not from
the merger of spiral-like galaxies. This is fully consistent with
the CG environment where early-type galaxies constitute the
dominant population \citep{coziol04}. Finding
evidence of dry mergers in CGs is important, because it suggests that 
these systems are obviously not in a dynamically stable state. It
also suggests that since the merging process of the galaxies is
not yet complete, these systems cannot be as old as that believed based
on the absence of standard observational merging evidence, like
luminous active galactic nuclei (AGNs), recent star formation events, or 
post-burst stellar populations in evolved galaxies.

What is missing in CGs is a time scale for the evolution process of
the galaxies. Was the evolution of galaxies accelerated in the group
environment? Is the dry merger the result of such evolution? Is the dry
merger limited only to dense environments? What is the role of the
potential of the group in the disappearance of the gas? How fast was the gas
exhausted or consumed? Did it burn rapidly forming the
bulges of numerous early-type galaxies, was it lost feeding a
black hole, or was it mostly ripped off of the galaxies and
lost to the intergalactic medium?

In order to find some answers to the above questions, we have decided 
important to extend our study of characteristic isophotal parameters
and asymmetry to two different structures having lower spatial
density than CGs: isolated galaxies and isolated pairs of galaxies.
Isolated galaxies are considered to have a low probability of
interaction with another galaxy of similar mass over a Hubble time 
\citep{vettolani86}. Consequently, a sample of isolated galaxies
can be treated as a set of ``comparison objects'', free during most
of their lifetime from environmental effects. Isolated galaxies are
uncommon in the universe where most of the galaxies tend to be
clustered in groups, as shown by Tully (1987). The reason for their
existence, therefore, may be an interesting subject of study by
itself. Isolated pairs of galaxies ar in the next level of galaxy
density. In the nearby universe, these systems are also rare and 
consequently their history is not well documented. Many questions
still need to be answered. How long have these galaxies been
interacting? Are they engaged in first encounters or did they
interact multiple times before with their companions? Are these
transient phenomena (high-energy orbits) or merging encounters (
low-energy or decaying orbits)?

For our study, we used three well-defined environment samples: the
Catalog of Isolated Galaxies (KIGs), from Karachentseva (1973), the
list of Isolated Pairs of Galaxies (KPGs), as compiled by
Karachentsev (1972), and the Catalog of Compact Groups of Galaxies
(HCGs) from Hickson (1982). Our analysis is based on the application
of two independent methods: the fitting of elliptical ellipses on
the isophotal levels of the galaxies and the determination of their
asymmetry \citep{cp07}. We present the
characteristics of the observed samples in Section~2 and describe
our observations and the reduction process in Section~3. In
Section~4, we explain the methods used for our analysis. The surface
photometry profiles, the color maps and asymmetrical images of the
observed galaxies, and the results of nonparametric statistical
tests used to establish the level of significance of the differences
observed are presented in Section~5. Finally, we discuss our results
and give our conclusions in Section~6. Our analysis for the near-infrared 
will follow in an accompanying paper \citep{plauchu10}.

\section{Selection and properties of the observed galaxies}

\subsection{Isolated Galaxies}

In 1973, Karachentseva used a simple method for identifying isolated
galaxies. By inspecting the blue prints of the Palomar Observatory
Sky Survey (POSS) she selected all the galaxies in the Zwicky
catalog \citep{zwicky61} whose nearest neighbor has a
size within a factor 4 of the major-axis diameter of the target
galaxy and lies more than 20 diameters away from it. This
definition implies that a galaxy with a diameter of 20 kpc and
peculiar velocity on the order of 150 km s$^{-1}$ has not been
influenced by a similar mass galaxy during the last $\sim 3$ Gyr 
\citep{turner79}. The KIGs are observed in the nearby universe
($z < 0.14$ with median $z = 0.02$) and have apparent magnitudes
brighter than 15.7, which is the limit of the Zwicky catalog 
\citep{zwicky61}. Members in this catalog have north
declination $\delta\ge-3\,^{\circ} $, the majority being at high
galactic latitudes ($\mid b \mid \ge20^\circ$).

\subsection{Isolated pairs of galaxies}

In the early 1970s, Karachentsev (1972) compiled what was at that time
the first list of pairs of galaxies, the Catalog of Isolated Pairs
of Galaxies (KPGs). Using the Zwicky catalog, the KPGs were selected
from visual inspection of the POSS prints based solely on their
observed properties, like apparent separation, apparent magnitudes,
and angular diameters, and without reference to apparent signs of
interaction. The pairs of galaxies in the KPG are also located in
the nearby universe ($z < 0.06$ with median $z$ = 0.02), all have
north declination $\ge-3\,^{\circ}$, high galactic latitude ($\mid b
\mid \ge20^\circ$), and photographic magnitudes brighter than 15.7.
This catalog is considered suitable for studying galaxies in pairs
because of its size, completeness, and relatively unbiased selection 
\citep{toledo99}.

Subsequent spectroscopic observations revealed that only half of the
KPGs in the initial catalog have small relative velocities,
$\Delta v <$ 100 km s$^{-1}$, while the remaining pairs have
relative velocities extending upward and as far as 10,000 s$^{-1}$ 
\citep{karachentsev87}. Later on various attempts
to establish solid criteria to determine true pairs based on the
relative velocity of the member galaxies were made. For example, Turner (1976)
proposed that physical pairs must have $\Delta v < 425$ km
s$^{-1}$. According to Makino \& Hut (1997), pairs of galaxies have
a higher probability to show effects due to interaction when the
difference in radial velocity between the two galaxies is comparable
or lower than their internal velocity dispersion. In the same vein,
Patton et al. (2000) suggested $\Delta v \le 500$ km s$^{-1}$. For
our sample we have followed the latter authors and selected pairs
with $\Delta v \le 500$ km s$^{-1}$.

\begin{figure*}
\epsscale{2.0} \plotone{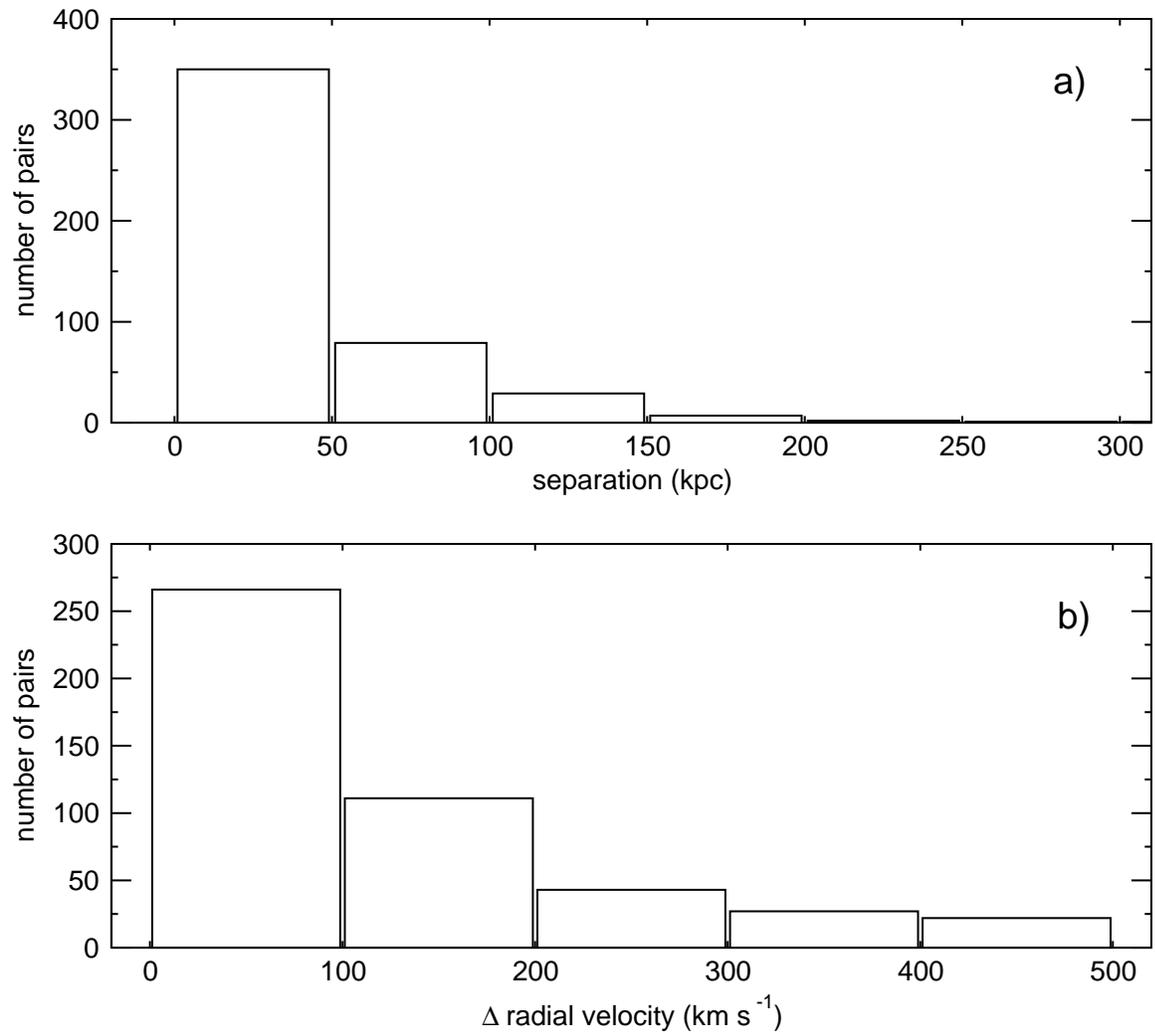}
\caption{ Distribution of pairs
as a function of (a) linear separation and (b) difference in radial
velocity of the pair members. \label{proppair}}
\end{figure*}

In Figure~\ref{proppair}, we show the linear separation and
difference in radial velocity between the members of the KPG pairs.
The majority are close pairs of galaxies, with spatial separation
lower than 50 kpc and difference in radial velocity lower than or equal
to 150 km s$^{-1}$. For comparison, our Local Group of galaxies as
viewed at a comparable redshift ($z \sim 0.02$ ) would look like a
pair with a spatial separation of ~772 kpc \citep{ribas05} and a
difference in radial velocity equal to 112 km s$^{-1}$. Therefore,
the galaxies in the KPG are much closer than the two major galaxies
in our Local Group.

\subsection{Compact Groups of Galaxies}

In the early 1980s, Paul Hickson conducted a visual search for CGs
using red POSS prints in order to obtain a homogeneous sample that
could be subjected to statistical analysis. Hickson's Compact
Groups Catalog forms one of the most studied samples to date 
\citep{hickson82}. The HCGs are small systems of three to eight galaxies in
apparent close proximity in the sky. The space density of galaxies
is very high, often exceeding that of the cores of large clusters of
galaxies \citep{hickson92}. The relatively low velocity
dispersions, which are generally comparable to galaxy rotation
velocities, make interactions and mergers common in these systems 
\citep{hickson92}. Many galaxies in the HCG show morphological
peculiarities indicative of gravitational interactions 
\citep{mendes94, cp07}.

In 1992, Hickson et al. (1992) obtained spectroscopic observations for
almost all the galaxies in the HCG (462 galaxies) and found that
only 92 groups are real bounded systems with at least three members 
and with a median radial velocity dispersion of 200 km s$^{-1}$. 
The HCGs are nearby universe structures (with $z <$ 0.14
and median redshift $z$ = 0.03), which are located well beyond the
Virgo Cluster \citep{hickson92}. For our study, we have selected
groups from Hickson et al. (1992) with velocity dispersions
$\sigma_{v} \le$ 800 km s$^{-1}$.

\subsection{Observed samples}

\begin{figure*}
\epsscale{2.0} \plotone{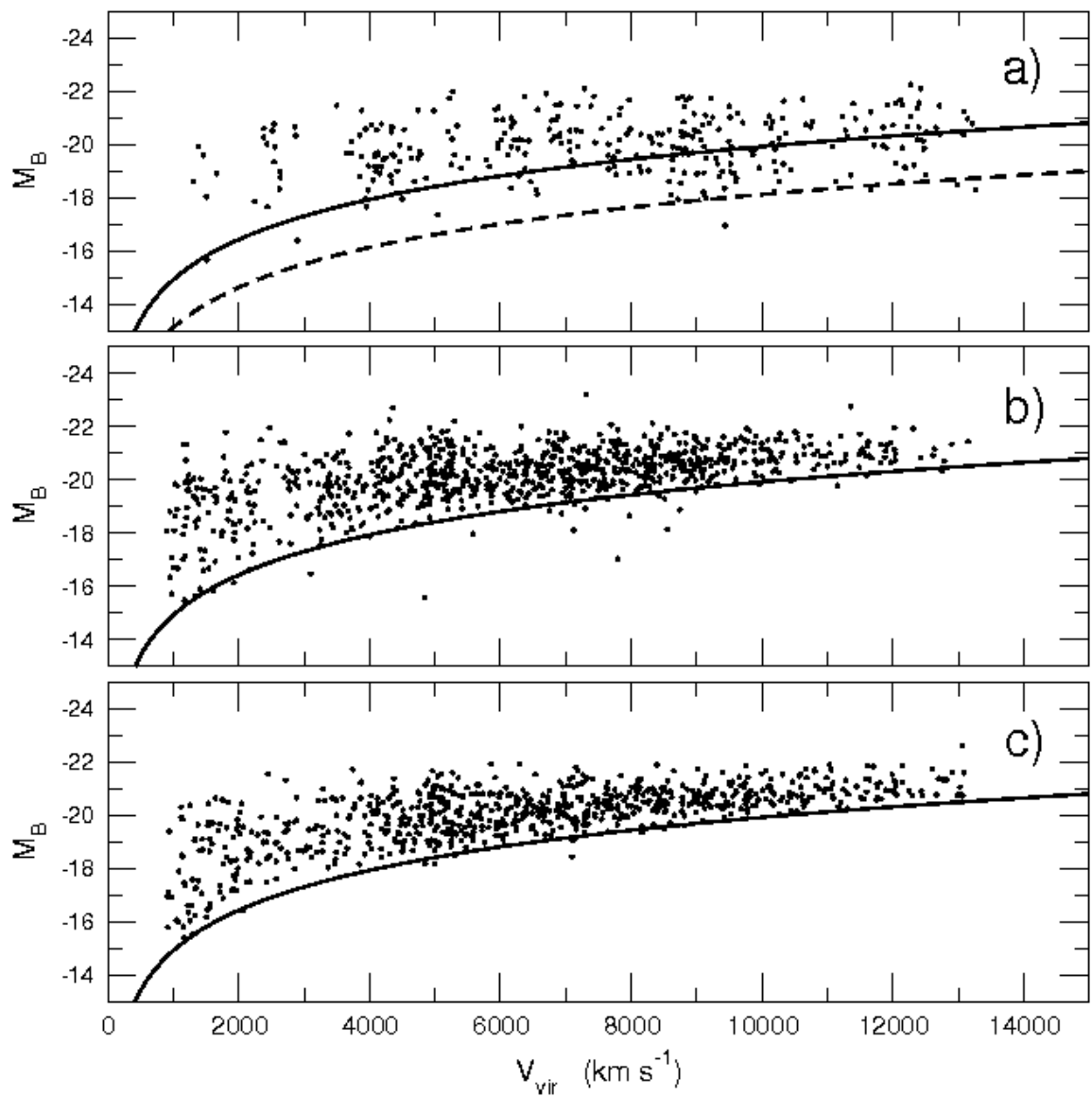} \caption{Distribution of absolute $B$
magnitude vs. virgocentric velocity for (a) HCG, (b)
KPG, (c) and KIG. The curves indicate the magnitude limit of the
survey for the HCG catalog $m_B$=17.5 ({\it dashed curve}) and the
KPG and KIG catalogs $m_B$=15.7 (continuous curve). The data
have been extracted from HyperLEDA. \label{magBcats}}
\end{figure*}

In Figure~\ref{magBcats}, we show the distribution of absolute $B$
magnitude versus the virgocentric velocity for 308 HCG galaxies
(top), 938 KPG galaxies (middle), and 777 KIG galaxies
(bottom). Only galaxies inside the range 900 km s$^{-1} \le v _{vir} \le$ 
13,000 km s$^{-1}$, with $M_{B}\le -15$, and satisfying
the previous selection criteria, are plotted in this figure. One can
see that the KPG and KIG surveys scan comparable volumes. The HCG
survey, on the other hand, being slightly deeper, contains
galaxies with lower luminosity ($M_{B}\le -18$) above $v_{vir}=
8000$ km s$^{-1}$. This difference will be taken into account during
our analysis.

Based on the samples we set up our targets on galaxies with
declinations in the range -32$\,^{\circ} \le \delta \le$ +55 $\,^{\circ}$ and
the semi-major axis in the range $0'.5 \le a \le  3'.5$, which allows for
optimal spatial resolution. Also, to minimize inclination
corrections for photometric data to evaluate of basic
structural parameters, we have applied an ultimate criterion based
on the semi-minor to semi-major axis ratio, keeping galaxies with $
b/a \ge 0.4$ (or $i\le70 \,^{\circ}$). Only in a few cases, in the
HCG, applying this last criterion was impossible.

Our final selection for the observed galaxies also depended on the
time allocated for observation and the weather conditions.
We were able to observe in total 214 galaxies: 37 KIGs, 71 KPGs, and 106
HCGs. All the galaxies have redshifts $z<0.04$. The properties of
these galaxies are reported in Tables~\ref{tabl1}-{tabl3}. For each of the galaxies,
we have double checked the morphological type. Most of the KIG
galaxies already had their morphology determined by Sulentic et al.
(2006). In the cases where our CCD images and isophotal study
suggested a bar, we have added this information to their
morphological description.

\begin{figure*}
\epsscale{1.0} \plotone{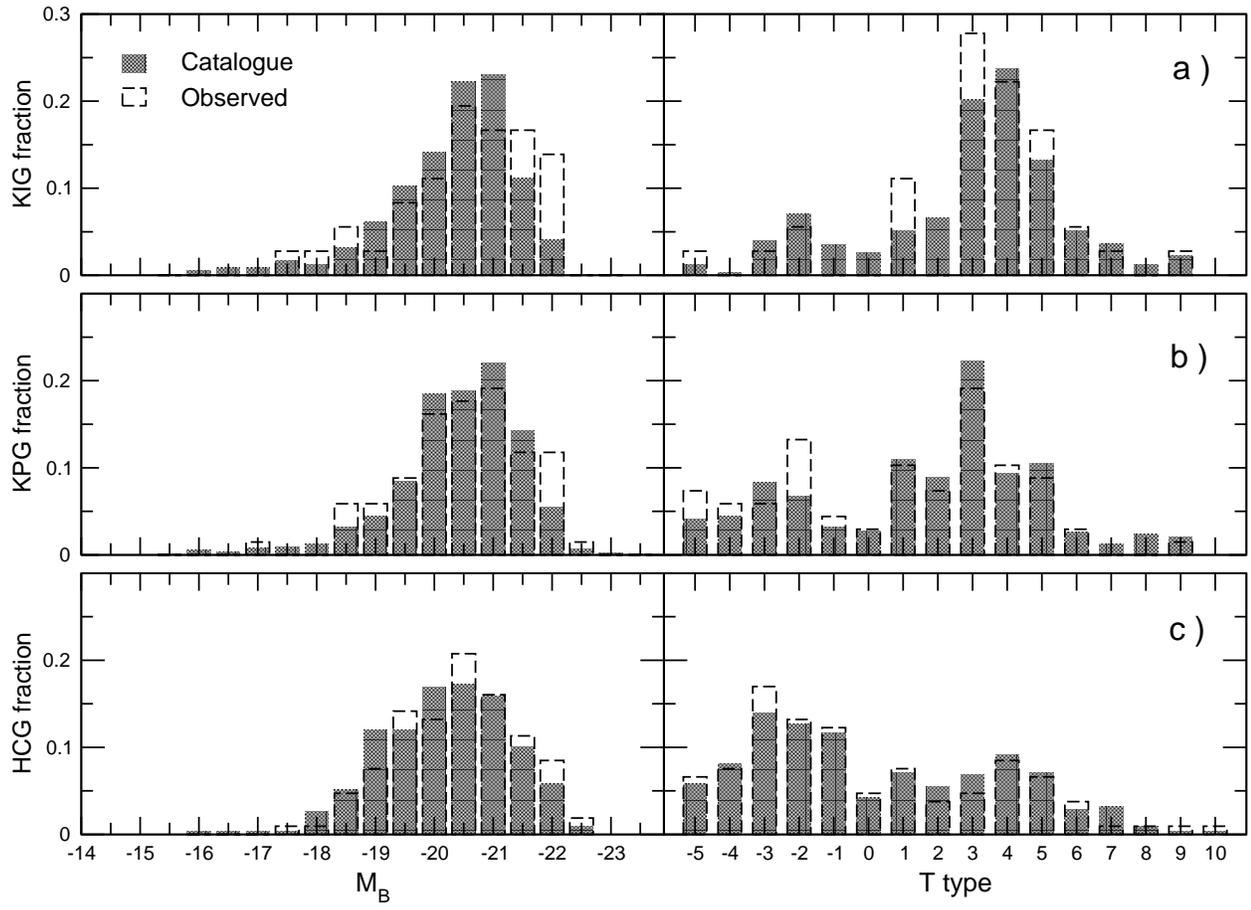} \caption{ Distribution of catalog
vs. observed galaxies (a) KIGs, (b) KPGs, and (c) HCGs. \label{comp}}
\end{figure*}

In Figure~\ref{comp}, we compare the characteristics of the observed
samples with the characteristics of the galaxies in their respective
catalogs. One can see that the observed samples reproduce the
absolute and morphological distribution of their parent samples
relatively well. The results of nonparametric statistical tests
(Mann-Whitney), presented in Table~\ref{tabl4}, are consistent with
no differences in absolute magnitude and size, although there seems
 a slight tendency for the observed KIGs to be nearer than the
galaxies in their parent sample.

\section{Observation and reduction}

The sample of 214 galaxies was imaged during five different
observing missions (see Table~\ref{tabl5}). The observations were
carried out using the 1.5 m telescope of the Observatorio
Astron\'omico Nacional, located at the Sierra San Pedro M\'artir in
Baja California, M\'exico. Depending on the observing run, there
were two different detectors attached to the telescope (see
Table~\ref{tabl5}): the Site1 and the Marconi CCDs. The first CCD
covers an area of about 4.3$^{'}\times4.3^{'}$\ on the sky, with a
spatial resolution of $0.26^{''}$\ per pixel. The second one covers
an area about 4.5$^{'}\times4.5^{'}$\, with a spatial resolution of
$0.28^{''}$\ per pixel, using a 2$\times$2 binning mode. For each
galaxy we took three images in each filter, with exposure times of
600-900 s in $V$ and 200-300 s in $I$. Each night
standard stars were also observed to calibrate the data in flux.
These stars were taken from the Landolt Equatorial Stars list 
\citep{landolt92} and cover a wide range in color: -0.30 $\le (V-I)
\le$ 2.63 or -1.12 $\le (B-V) \le$ 2.33.

The nights were clear during the last four observing runs, with
average effective seeing conditions at the telescope of $1.6^{''}$,
$1.9^{''}$, $1.8^{''}$, and $1.4^{''}$, respectively. During the
first run, all nights were not totally clear, with an average
effective seeing of $2.4^{''}$. During this run, we used a binning
mode of 2$\times$2 with the Site1 CCD to increase the signal-to-noise (S/N).
Note that because the surface brightness, ellipticity, position angle (P.A.), and
asymmetry profiles depend only on the S/N and spatial resolution in
a frame, a high photometry accuracy is not important for our
analysis.

The data were reduced and calibrated using standard algorithms in
IRAF\footnote{ IRAF is the Image  Analysis and Reduction Facility
made available to the astronomical community by the National Optical
Astronomy Observatory, which is operated by AURA, Inc., under
contract with the U.S. National Science Foundation.}. The images
were first trimmed in order to remove bad lines and columns at the
edges from the CCD and to reduce vignetting effects. We subsequently
applied a mask on all the images to remove the bad pixels on the
CCD. An average bias (combining 15-20 bias images) was subtracted
from the object images and the flat frames. Several sky flat frames
taken in each filter at the beginning and/or end of each night were
normalized, combined, and then divided from each object image. The
level of un-flattening is well below 2\%, and the flat fielding
conserves the flux to better than 99\%. Cosmic-ray removing was done
using the \emph{COSMICRAYS} task. Resilient cosmic rays were
corrected by hand using the \emph{IMEDIT} task. To each image, a
small shift (a tenth of a pixel) was applied to correct telescope
drifting or repositioning. After trimming the images to the same
dimension, they were averaged together. The final reduction step
consisted of eliminating the sky contribution. This was done by
measuring the mean flux within 5$\times$5 pixel boxes all around the
targets (where there are no stars or background objects) and
subtracting this value from the images.

The instrumental magnitudes were estimated by measuring the flux of
each observed standard star. Air-mass correction was applied using
extinction coefficients proper to San Pedro M\'artir \citep{schuster01}. 
The calibration equations were determined by fitting
linear regressions on the observed values. For photometric errors, we
adopt the standard deviation between our estimated magnitude and the
magnitude determined by Landolt (1992). Magnitudes for the observed
galaxies have also been corrected for galactic extinction \citep{schlegel98}. 
Due to the low redshifts of the galaxies (median
$z=0.02$) no $K$-correction was applied, since these are smaller
(e.g., 0.02 in $V-I$) than our uncertainties. The general
characteristics of the observations are given in Table~\ref{tabl5}.

Note that the calibration in flux was done after the different
analysis (ellipse fitting and asymmetry) were performed. This method
avoids keeping the noise in the images at low level producing the
highest possible S/N.

\section{Description of the analysis methods}

Three analyses were performed on each galaxy: fitting of
ellipses, formation of color maps, and estimation of asymmetry
level. Here we describe the methods used and information obtained
from each analysis.

Our analyses in different bands (optical and near-infrared, (NIR)) yield information
over different time-scales. In the optical, our analysis is sensible
to young or intermediate age stellar populations and dust
extinction. In the NIR, our analysis is sensible to
older stellar populations, and consequently to mass distributions 
\citep{cp07, plauchu10}.

\subsection{Isophotal method analysis and color maps}

Surface photometry was carried out on each galaxy. This was done
within STSDAS\footnote{STSDAS is distributed by the Space Telescope
Science Institute, which is operated by the Association of
Universities for Research in Astronomy (AURA), Inc., under NASA
contract NAS 5-26555} with the
\begin{footnotesize}ELLIPSE\end{footnotesize} task \citep{jedrzejewski87}. 
The algorithm used by this task derives various geometric
parameters, such as surface brightness $\mu$, ellipticity
$\epsilon$, P.A., and the harmonic amplitude $B_{4}$.
This last parameter is related to the standard fourth-order Fourier
cosine coefficient, $a_{4}$ \citep{bender88}, normalized to the
semi-major axis $a$ at which the ellipse was fitted ($a_{4}/a =
B_{4}\sqrt{1-\epsilon}$). Another important parameter is the $V-I$
color index profile. This profile is obtained by measuring the $V$
and $I$ magnitude profiles, subtracting one from the other.

\begin{figure*}
\epsscale{0.8} \plotone{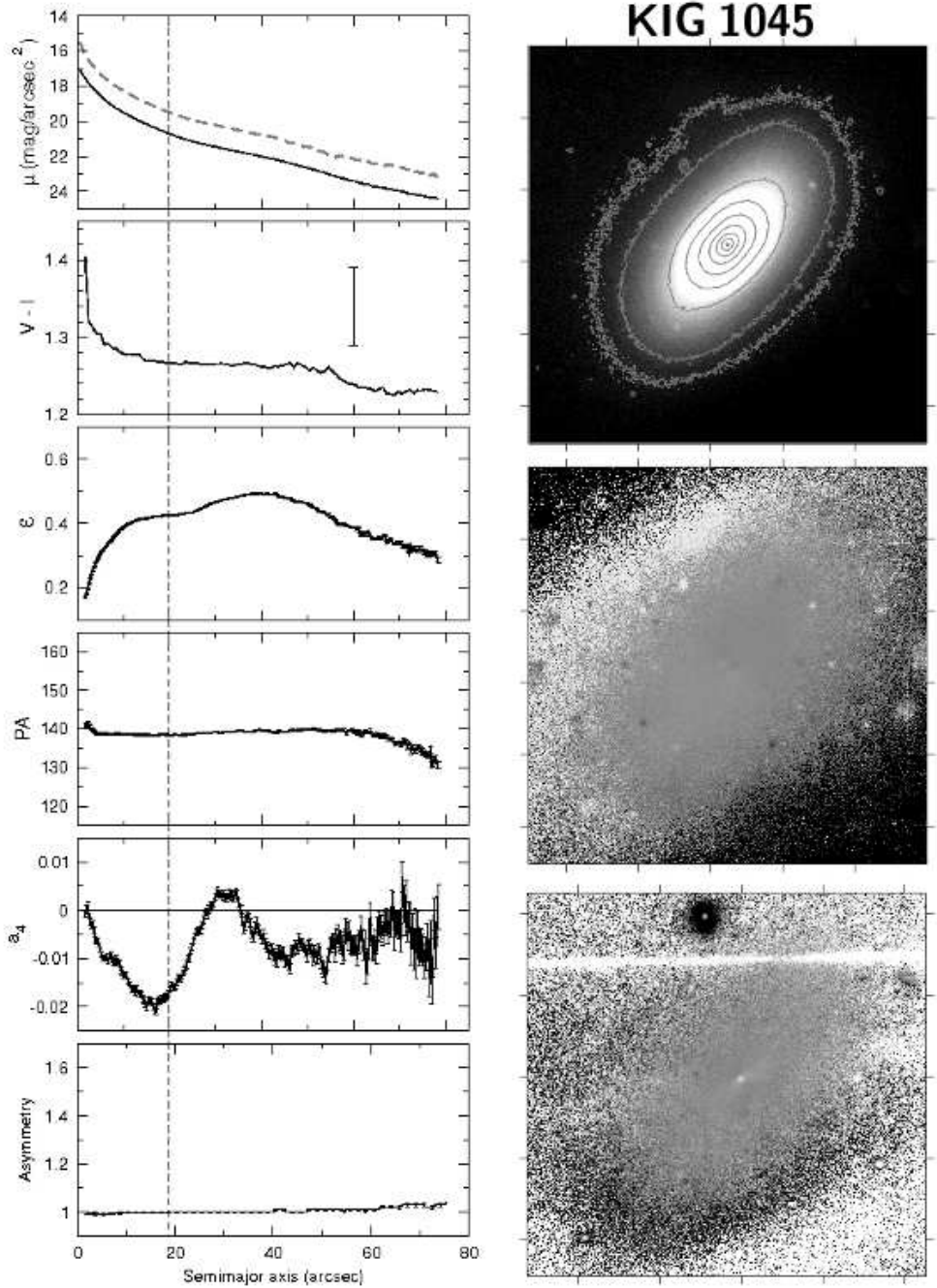} \caption{ KIG 1045 mosaic. Left
panel: isophotal parameter profiles as a function of semi-major
axis $a$  $-$from top to bottom: surface brightness $V$ (solid curve)
and $I$ (dashed curve), $V-I$ index color, ellipticity, P.A., 
isophotal deviation from pure ellipse, and asymmetry level.
The dashed vertical line indicates the average half-radius $r_0 = 5$
kpc (or $r_0 = 2.5$ kpc when indicated). Right panel: $V$-band
image, displayed on a logarithmic scale with superimposed isophotes;
the residual image (middle right) and the $V-I$ color map
(bottom image). North is up and east is to the left. (The complete 
figure set (214 images) is available in the online journal.)
\label{mosaicOPT}}
\end{figure*}

The only requirement for
\begin{footnotesize}ELLIPSE\end{footnotesize} to work is an initial
guess of the geometric center, $\epsilon$, and of P.A., of the galaxy. 
The geometrical center of a galaxy is
determined by locating the peak in light distribution \citep{cp07}. 
The default $\epsilon$ and P.A. were 0.05 and 0, respectively. The task 
\begin{footnotesize}ELLIPSE\end{footnotesize} is applied keeping the
geometric center fixed and allowing $\epsilon$ and P.A. to vary.
This method yields surface brightness and color profiles that match
the local variations of the structural components. We also restrict
the fit of ellipses in the central part of the galaxy to a radius
larger than the seeing (see Figure~\ref{mosaicOPT}) and minimize the
light contribution from companion galaxies (important only for the
HCGs and KPGs) by stopping the task manually at the maximum radius
possible.

The isophotal parameters measured by
\begin{footnotesize}ELLIPSE\end{footnotesize} provide important
information on the physical morphology and are intimately related to
the dynamical properties of the galaxies \citep{barth95, cp07}.  
For example, large variations in 
P.A.$\sim$20$^\circ$, or twists \citep{nieto92}, usually reflect
inhomogeneous mass distributions \citep{zaritsky86}, while
$\epsilon$ variations reflect bars, dust, mass perturbations, or
small disks in the central regions of early-type galaxies. Early
isophotal studies have shown that large isophotal twists are only
measured in interacting galaxies, suggesting that they are the
consequence of close encounters or mergers \citep{kormendy82, bender87}.  
In our analysis we define a twist, $\theta$, as
a variation in P.A. accompanied by a monotonically varying
ellipticity $\epsilon$ with amplitude $\Delta \epsilon \ge 0.1$.
This definition allows differentiating between variations of P.A.
produced by triaxiality and from those produced by interactions 
\citep{kormendy82, bender87}.

The isophotal deviations from the pure ellipse, quantified by the
$a_{4}$ coefficient, determine the characteristic shape of the
isophote: boxy ($a_{4}<0$), consistent with a pure elliptical or round
isophote ($a_{4}\sim0$), or disky ($a_{4}>0$), consistent with a
slightly pointed isophote. Elliptical galaxies with disky isophotes
tend to be faint. They generally contain a rapidly rotating stellar
population with a nearly isotropic velocity dispersion. Elliptical
galaxies with boxy isophotes tend to be luminous and massive. They
have anisotropic velocity dispersion and are thought to be supported
by pressure rather than stellar rotation. These differences suggest
two distinct formation scenarios for boxy and disky elliptical
galaxies. Numerical simulations have shown that the formation
processes depend highly on the initial conditions: initial mass
ratios, individual angular momentum, dust and gas contents of the
merging galaxies \citep{hernquist93, barnes96}. The general idea is
that mergers between unequal-mass, gas-rich galaxies tend to produce
disky early-type remnants, while mergers with equal-mass,
high-density, and gas-poor galaxies tend to produce boxy remnants 
\citep{naab03}.

The concentration index, $C$, is a measure of the light concentration
of a galaxy profile, having high values for centrally concentrated
light profiles. It is well know that the concentration index has a
tight correlation with morphological type \citep{abraham94, shimasaku01}; 
early-type galaxies tend to have the most
concentrated light profiles, while late-type galaxies have the least
concentrated ones. Interactions between galaxies can also perturb
the stellar material, changing the light profiles of
the galaxies in the process and affecting their concentration index.

Because we are studying nearby galaxies, the spatial resolution of
our CCD images allows us to measure the $C$ parameter at different
radii. Based on the isophotal profiles of the galaxies, we have
estimated $C$ values inside and outside a physical radius $r_{0}$,
independent of the distribution of light. To estimate this radius,
we have used the major axis at 25 mag arcsec$^{-2}$ \citep{paturel03} 
as given in Hyperleda and determined the linear diameters in
$B$ magnitude, $D_{B}$, for the galaxies in the HCG, KPG, and KIG
catalogs estimating the median of the three distributions. The
median value obtained is 23 kpc. In our sample, a few galaxies (20\%
of the sample: 26 HCGs, 12 KPGs, and 4 KIGs) turned out to have a
$D_{B}$ that is smaller than this value. Consequently, we have used
two different $r_{0}$; one is equal to 5 kpc (approximately $D_{B}/4$)
for the standard size galaxies (approximately $M_V < -20$) and the
other is half of this value, 2.5 kpc, for galaxies with smaller diameters. For our
analysis we give three concentration indices: one inside $ r_{0}$,
$C(r< r_{0})= \mu (r=r_{0})$--$\mu (r< r_{0})$, one outside $
r_{0}$, $C(r> r_{0})=\mu (r> r_{0})-\mu (r= r_{0})$ and a global (or
total) concentration index $C_{Total}=\mu (r> r_{0})-\mu (r<
r_{0})$.

For the sake of comparison, in our analysis we have also choosen a radius
which depends on the light distribution in the galaxies. We have
used the Petrosian, $R_P$, and effective radii, $R_e$, as determined
based on a modified form of the Petrosian (1976) system \citep{graham05}. 
In this system $R_P$\ is defined as the projected radius
where $1/\eta (R_P)=0.2$. The Petrosian index, $\eta (R) = \langle
I\rangle_R/I(R)$, is the ratio of the intensity of an isophote at
radius $R$\ and the average intensity within that radius, as
measured using circular apertures. In the case of small galaxies,
where the faint central surface brightness does not allow us to reach
$1/\eta (R_P)=0.2$, we have used $1/\eta (R_P)=0.4$\ instead. In
Tables~\ref{tabl6} and \ref{tabl7}, for all the galaxies in
our analysis we give $R_P$, $R_e$, the magnitude inside the effective
radius; $M_e$, a concentration index, which is defined as the ratio
of radii that contain 90\% and 50\% of the Petrosian flux,
$R_{90\%}/R_{50\%}$; the surface brightness at the effective radius,
$\mu_e$; and the color at this radius.

Finally, as a complementary analysis, we have constructed a $V-I$
color map for each observed galaxy. We present these maps in the
bottom right part of Figure~\ref{mosaicOPT}. In these maps, bright
regions are consistent with red stellar population or dust
extinction, and dark regions are consistent with blue stellar
populations. These color distribution maps were found to be
extremely useful in detecting tidal tails, galaxy satellites, dusty
patches, and common envelopes in galaxies.

\subsection{Asymmetry method analysis}

Another useful method for the study of morphology consists of
estimating the level of asymmetry of a galaxy \citep{abraham94, abraham96, 
schade95, conselice97, conselice00, cp07, hutchings08}. The 
leitmotif behind this method is that the level of asymmetry of a
galaxy reflects something about its history of formation and
evolution. For example, galaxies that are old and already well
evolved, or galaxies that formed in isolation, are expected to
possess fairly symmetric distributions of light. On the other hand,
galaxies affected by interactions or mergers sometime during their
evolution are expected to show more complex distributions.

The interpretation of asymmetry may seem straightforward enough for
early-type galaxies, but it is not that simple for later-type spiral
galaxies. Various studies have shown that in late-type spirals,
asymmetric structures may result from intrinsic processes related to
star formation \citep{schade95, conselice00}. Extra
care must be taken, therefore, before drawing clonclusions about the origin of
asymmetries in any sample of galaxies. In Coziol \& Plauchu-Frayn
(2007) it was shown that the asymmetric structure analysis is
complementary to the isophotal one: there is a one-to-one relation
between the variations of isophotal characteristic parameters and
the existence of asymmetries related to inhomogeneous distribution
of mass produced by interaction effects. Applying the two analyses
in parallel yields a high confidence level when interpreting the
results.

For the present analysis we have used a slightly different
measure of asymmetry than that found in the literature (also different from 
the one used in Coziol \& Plauchu-Frayn 2007). This was done in
order to make the interpretation more straightforward. The principle
of the asymmetry method is relatively simple (see Coziol \&
Plauchu-Frayn 2007 for details). The image of a galaxy is rotated by
$180^{\circ}$ and divided from the original image. Any differences
in the distribution of light (asymmetries) appear under the form of
excesses of light (bright regions), together with their
corresponding shadows (dark regions) on the opposite side (see
Figure~\ref{mosaicOPT}).

To measure the asymmetry level, the residual images are smoothed
using boxes of size equal to the seeing in pixels, reducing the
noise. Ellipses are then fitted to the residual image of each
galaxy, keeping the center, ellipticity, and P.A.fixed. 
The level of asymmetry as a function of semi-major axis 
$a$ is estimated by the following formula:

\begin{equation}
A(a)_{180^\circ}\equiv \frac{I_0}{I_{180^\circ}}
\end{equation}

where $I(a)_0$ is the intensity in the original image and
$I(a)_{180}$ is the intensity in the rotated image. This formula yields
values between 1 (completely symmetric) and $>1$\ (completely
asymmetric).

Comparing with the residual images ($I_0/I_{180}$) it is easy to
verify that asymmetries in our analysis appear as structures in the
asymmetry curve. The amplitudes of these structures are
proportional to their relative intensities. For example, an
asymmetry of $A=1.2$ indicates a concentration of light $20\%$
brighter than the intensity at that radius on the opposite side.
This corresponds to a high level of asymmetry. On the other hand, a
level of asymmetry of $A=1.0$ indicates that the intensity of light
is the same on both sides (complete symmetry). In our analysis, a
symmetric distribution of light yields a flat asymmetry curve (see
the bottom left graphic in Figure~\ref{mosaicOPT}).

For our asymmetry analysis determining the center of the galaxies
around which the rotation is done is a crucial step. If this is not
done carefully spurious asymmetries can be produced. The method we
used (finding the peak in luminosity; see Coziol \& Plauchu-Frayn
2007) is simple and yields excellent results. It has also the
advantage to correspond to the same center as used during the
isophotal analysis. As a check, one can verify that, as expected, at
the center of the galaxies the asymmetry curves have a level
of 1 (minimum asymmetry). Moreover, real asymmetries produce
isophotal structures that are detected by our first analysis based
on ellipses fitting. Therefore, we are secure that no spurious
asymmetries are produced by our method.

Our analysis is not sensitive to sky gradients, because it applies
to the inner part of the galaxies, minimizing the possible
contamination by foreground stars. When needed, foreground stars
were eliminated using masks (using IMEDIT in IRAF). When a star was
found lying very near the body of a galaxy, a special mask was used
within \begin{footnotesize}ELLIPSE\end{footnotesize} itself. In
cases where it was impossible to eliminate the contaminating star the
galaxy was rejected.

\section{Results of analysis}

In Figure~\ref{mosaicOPT}, we show the mosaic for one very symmetric
galaxy (the full sample of mosaic images is available in the online
version of the journal). On the left of the figure, we present the isophotal 
profiles, where the 
dashed vertical line indicates an average half-radius of $r_0=5$\
kpc (or $r_0=2.5$\ kpc, as used for small size galaxies). On the
right, we present the $V$-band image, displayed on a logarithmic
scale with superimposed isophotes. We also present the residual
image from the asymmetry analysis (middle right) and the $V-I$ color
map (bottom image). In all these images, the north is at the top and
east is to the left.

\begin{figure*}
\epsscale{1.1} \plotone{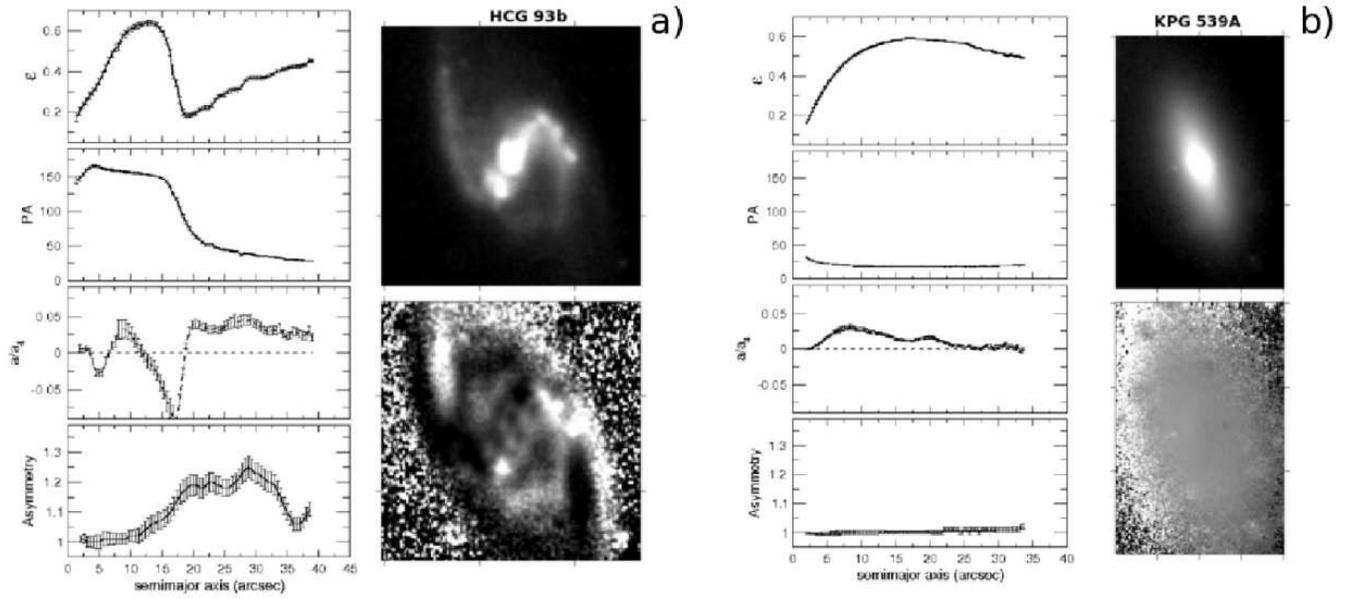} \caption{ Isophotal and asymmetry
profiles (a) the asymmetric galaxy, HCG~93b and (b) the symmetric
galaxy KPG~539A. The images show that both methods (isophotal and
asymmetry) are complementary: variations in isophotal parameters 
follow variation in asymmetry and vice versa. \label{isopasycomp}}
\end{figure*}

In Coziol and Plauchu-Frayn (2007) we have shown that the isophotal
and asymmetry analysis are consistent, yielding complementary
information. We will not repeat this analysis here, but give only 
two examples. In Figure~\ref{isopasycomp}(a), we show the asymmetrical
galaxy HCG~93b. The level of asymmetry increases by 20\% at a radius
of 18 arcsec. This asymmetry is accompanied by a sudden
significant variation in the three isophotal parameters. In the case
of the symmetric galaxy KPG~539A, Figure~\ref{isopasycomp}(b), the
absence of asymmetries is accompanied by a smooth variation in the
isophotal parameters.

\subsection{Comparison of galaxies with same morphologies in different environments}

For our analysis, we have divided our samples into three morphology
groups: early type (E$-$S0), intermediate type (Sa$-$Sb), and late types
(Sbc$-$Im). The median characteristics of the galaxies in these three
different groups are reported in Tables~\ref{tabl8}$-$\ref{tabl10}
for properties measured at radius $r_0$ and in Tables~\ref{tabl11}
$-$\ref{tabl13} for properties measured at radius  $R_e$. We now discuss the variations of
the various characteristics encountered in each group depending on
their environments. To check for the statistical significance of the
variations observed, nonparametrical tests (Kruskal$-$Wallis or
Mann$-$Whitney and Dunn's post-tests) were also performed. All the
tests were done at a level of significance of 95\%, which is the
standard for these kinds of tests. Description of the tests used can
be found in Coziol (2003). The results of the statistical tests are
reported in the last columns of Tables~\ref{tabl8}$-$\ref{tabl10}
and Tables~\ref{tabl11}$-$\ref{tabl13}.

\subsubsection{Early-type (E$-$S0) galaxies}

\begin{figure*}
\epsscale{0.7}
\plotone{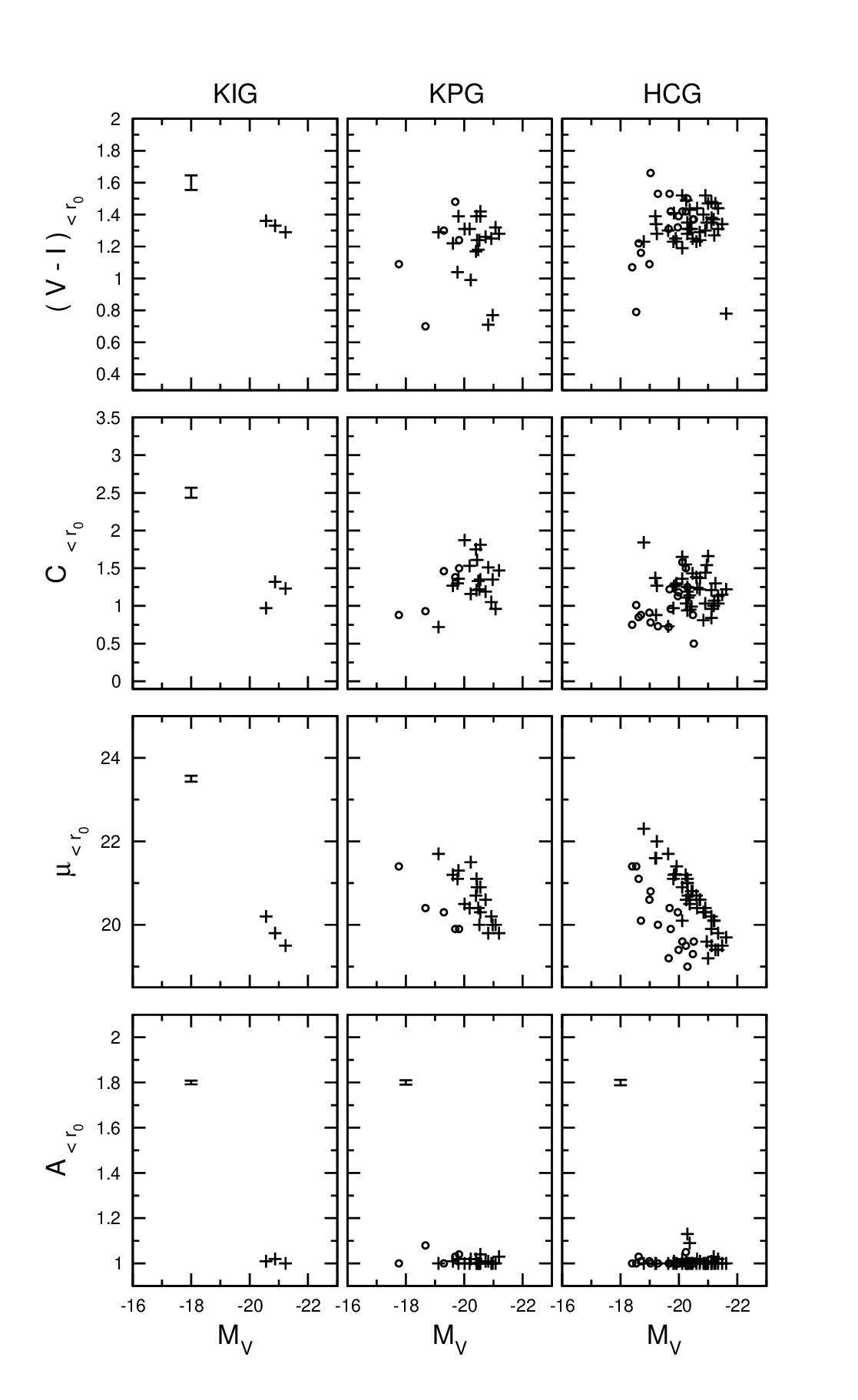}
\caption{Variations in arly-type galaxies of
isophotal parameters and asymmetry as a function of absolute magnitude in $V$ inside
$r_0$ for $-$from left to right: $-$KIG, KPG, and HCG early-type galaxies. The parameter
values are measured  inside the average half-radius $r_0 = 5$ kpc. For smaller size
galaxies (open circle), the average half-radius is $r_0 = 2.5$ kpc.
\label{tempIn}}
\end{figure*}

\begin{figure*}
\epsscale{0.7} \plotone{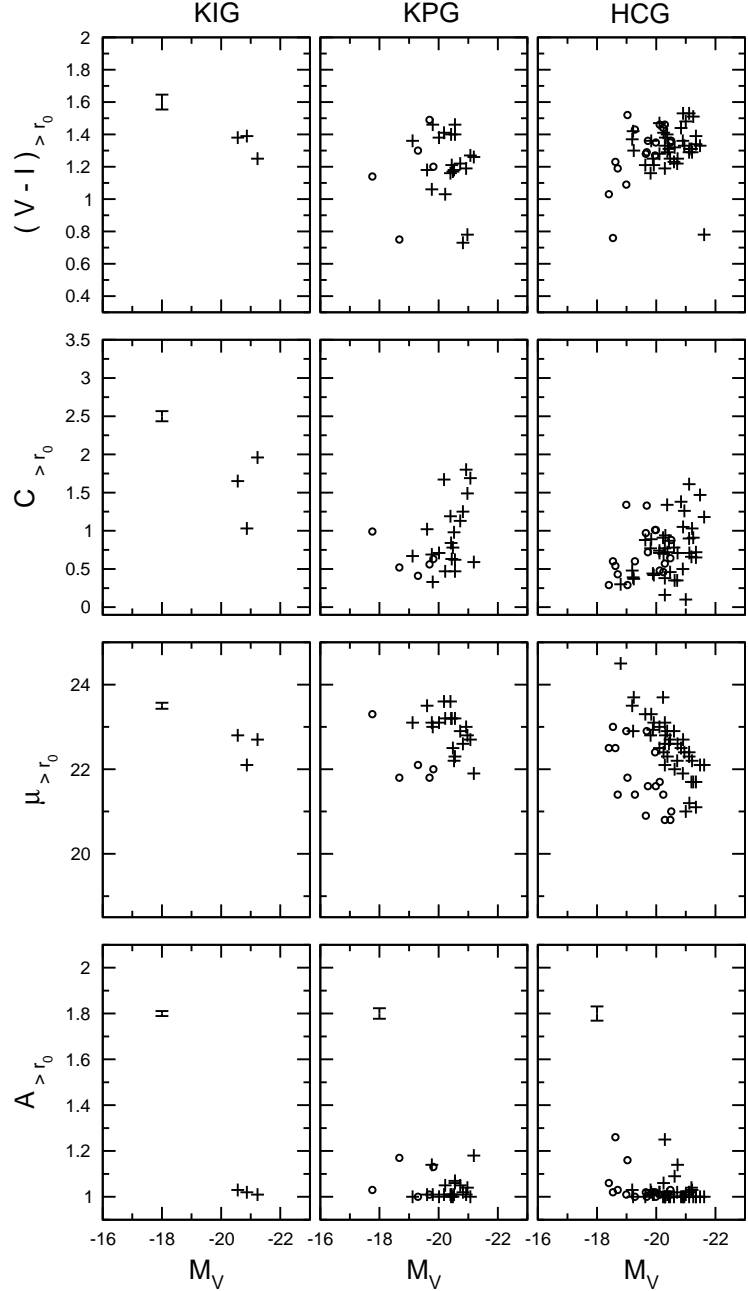} \caption{Variations in early-type
galaxies of isophotal parameters and asymmetry as a function of
absolute magnitude in $V$ inside $r_0$ for $-$from left to right: 
$-$KIG, KPG, and HCG early-type galaxies. The parameters are measured
outside the average half-radius $r_0 = 5$ kpc. For smaller size
galaxies (open circle), the average half-radius is $r_0 = 2.5$ kpc.
\label{tempOut}}
\end{figure*}

In Figures~\ref{tempIn} and \ref{tempOut}, we show the variations of
the isophotal parameters in early-type galaxies internal to $r_0$
(Figure~\ref{tempIn}) and external to $r_0$ (Figure~\ref{tempOut}).
In each graph, the {\it x}-axis represents the absolute magnitude in
$V$, as estimated inside $r_{0}$.

In this morphology group, there are only three galaxies that belong
to the KIGs. We have discarded these from our statistical tests. The
KPG galaxies in this morphology group tend to be slightly bluer than
the HCG, and this is independent of the radius and absolute
magnitude of the galaxies. This is confirmed by our statistical
tests (see Tables~\ref{tabl8} and \ref{tabl9} for $r_{0}$ and $R_e$,
respectively). Inside the half-radius, the HCG galaxies tend to be
less concentrated than the KPG galaxies. This is also confirmed by
our statistical tests (see Table~\ref{tabl8}).

The higher concentration and bluer color observed for the E$-$S0 KPG
galaxies are consistent with the idea of recent gas accretion and an
increase of star formation in the center of these galaxies. Outside
the half-radius, there are no differences in concentration between
the KPGs and the HCGs. This also agrees with the absence of difference
based on $R_{90\%}/R_{50\%}$, since this parameter is estimated at
comparable radii (Table~\ref{tabl9}).

There are no significative differences between the KPG and HCG
galaxies in surface brightness inside the half-radius. Father out
the HCG galaxies tend to have slightly higher surface brightness
than the KPGs (see Table~\ref{tabl8}). Since we are observing in the
optical, this suggests older stellar populations or more relaxed
structures as a whole in the HCGs, which is also consistent with the
slightly redder colors for the HCG galaxies.

Due to the low value of $R_e$ compared to $r_{0}$ , the statistical
tests find higher surface brightness on average for the early-type
KPG galaxies as compared to the HCGs (see Table~\ref{tabl9}). This
is consistent with our interpretation of more relaxed populations of
stars in the HCGs than in the KPGs.

In terms of asymmetry, we do not find any significative differences
among the samples. This morphological type appears to be very
symmetric independent of the environment. This suggests similar
formation mechanisms for these galaxies.

\begin{figure*}
\epsscale{0.8} \plotone{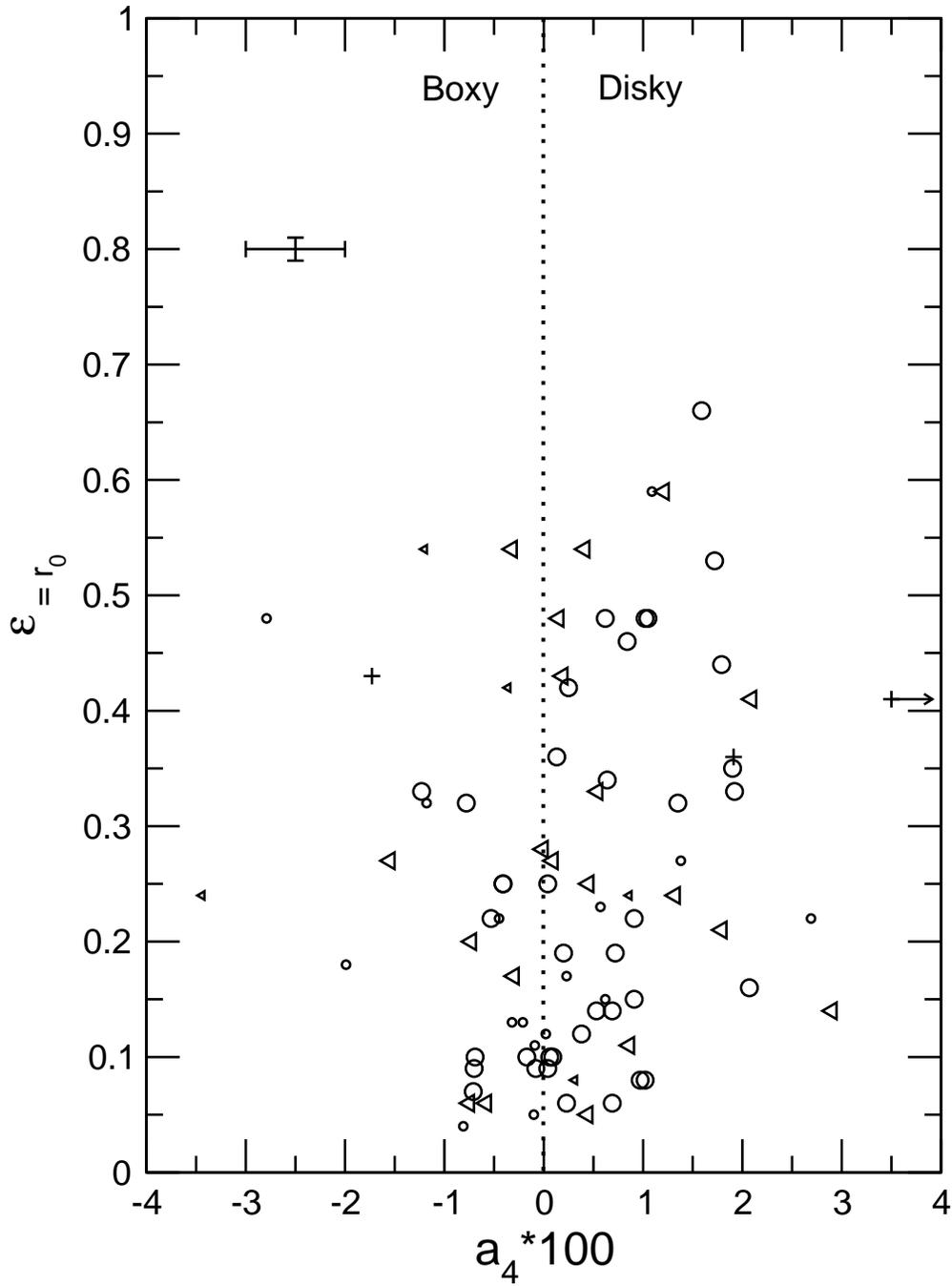} \caption{Isophotal shape as based on
the $a_{4}$ parameter vs. ellipticity, $\epsilon$, for early-type
galaxies. Both values were measured at $r_0$. Symbols correspond to HCG (
open circles), KPG (triangles), and KIG (plus signs)
galaxies. Smaller symbols correspond to small size galaxies (with
$r_0=2.5$ kpc). \label{boxky}}
\end{figure*}

For early-type galaxies, we can verify what types of isophotes are
prevailing: boxy with $a_{4}<0$ or disky with $a_{4}>0$. In
Figure~\ref{boxky} we show the values of $a_{4}$ as measured at the
half-radius for the E$-$S0 galaxies in different environments. One can
see that both the HCG and KPG E$-$S0 galaxies tend to occupy the
region of disky galaxies: 38 out of 57 (67\%) of the HCGs and 15 out
of 25 (60\%) for the KPGs. The ellipticity of these galaxies is also
quite high. This is consistent with the hypothesis of similar
mechanisms for the formation of these galaxies in both environments.
For example, the transformation of later-type spirals through gas
accretion and star formation in the central part would be one way to
produce E$-$S0-like galaxies that have disky rather than boxy
isophotes.

\begin{figure*}
\epsscale{0.7}
\plotone{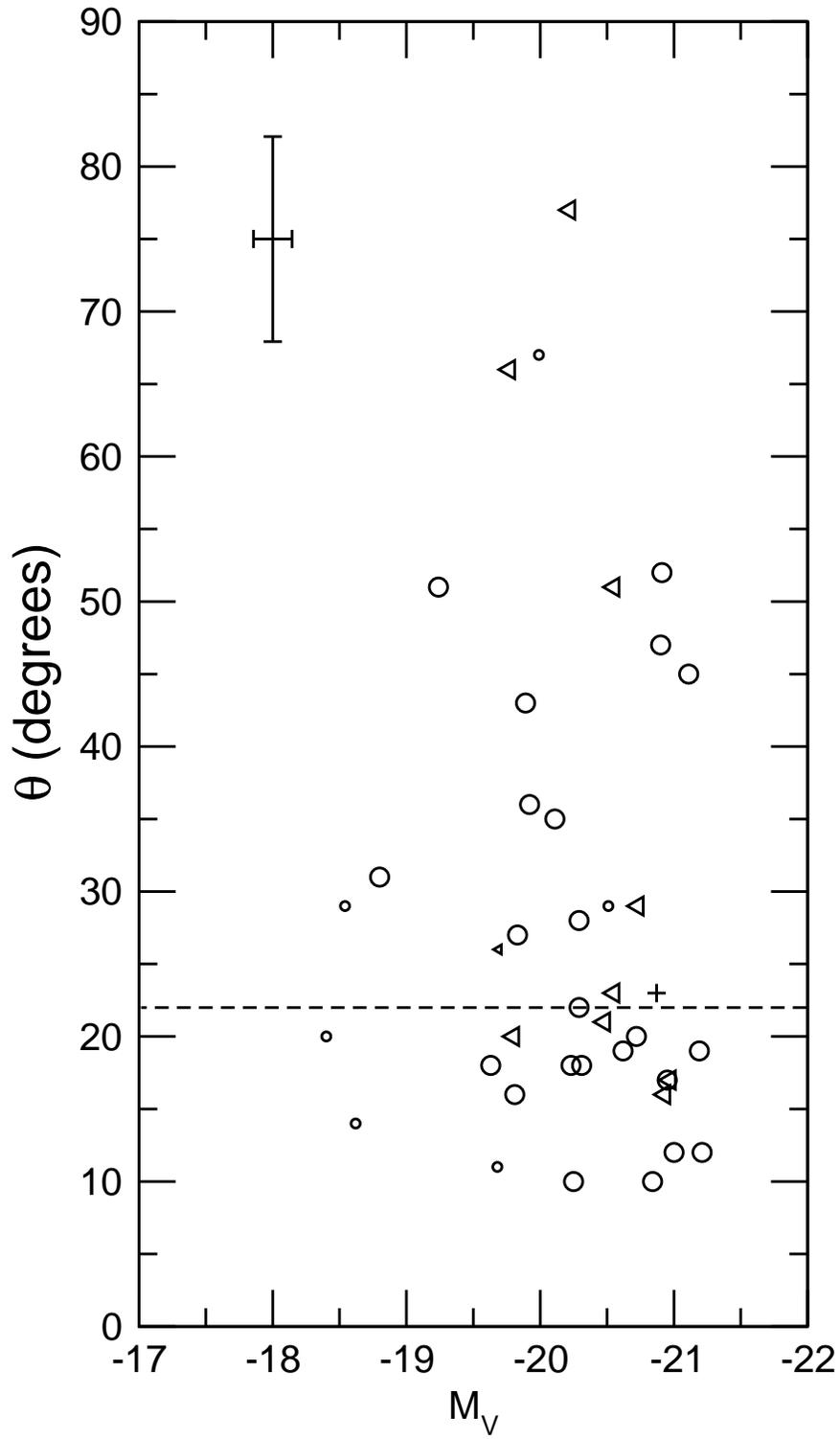}
\caption{Early-type galaxies with high twists, $\theta$,
as a function of absolute magnitude in $V$. Symbols are the same as in Figure~\ref{boxky}.
\label{twist}}
\end{figure*}

Evidence in favor of similar mechanisms for the formation of E$-$S0
galaxies in the KPGs and HCGs can also be found in the high frequency
of detection of isophotal twists in both samples: 40\% (10/25) in
the KPGs and 51\% (29/57) in the HCGs. The levels of the twists in
these galaxies are shown in Figure~\ref{twist} as a function of 
absolute magnitude in $V$. We consider large twists to be those with values
$\theta >$ 20$^\circ$. The median values of $\theta$ are 25$^\circ$
and 20$^\circ$ for the KPGs and HCGs, respectively.

\begin{figure*}
\epsscale{1.0} \plotone{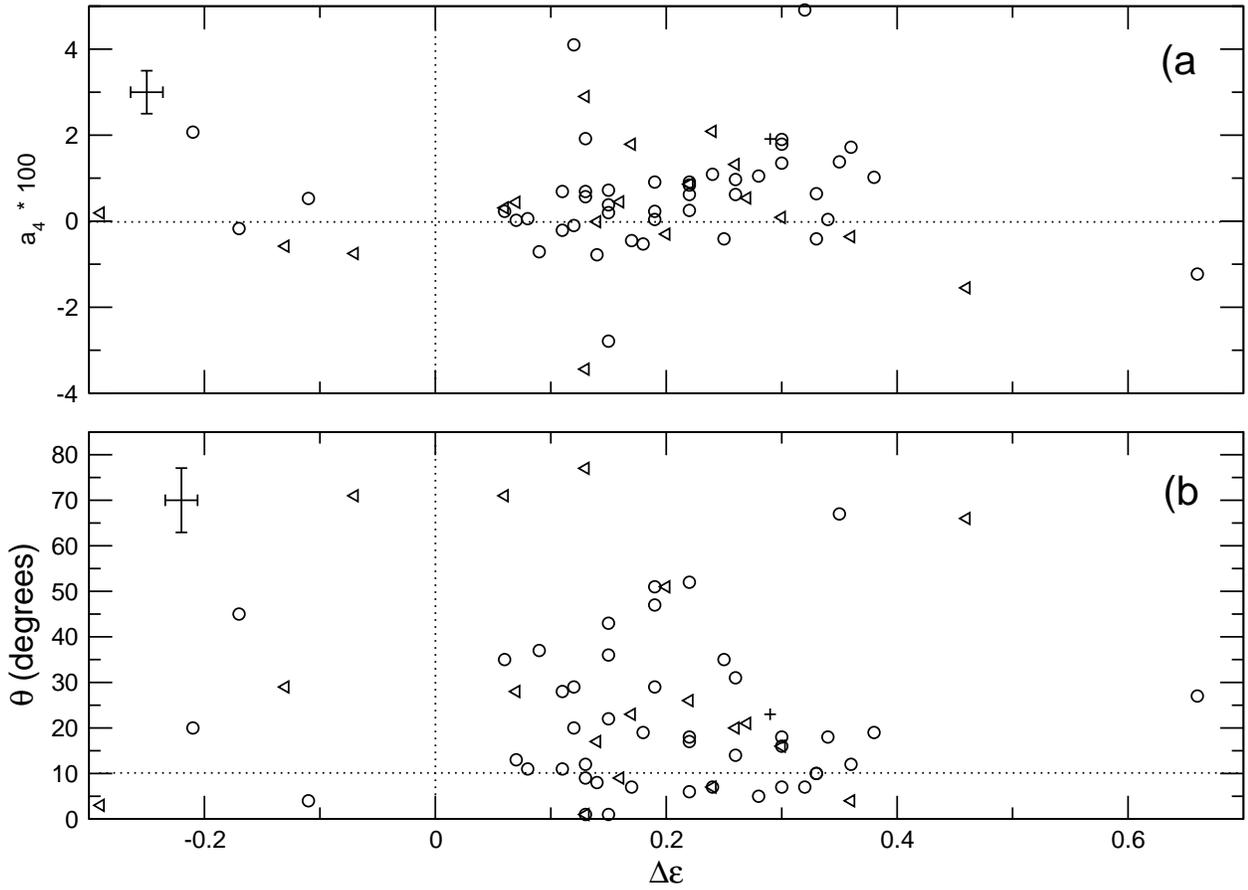} \caption{ Ellipticity variation,
$\Delta \epsilon$ vs. (a) isophotal shape $a_{4}$ and (b) twist
$\theta$ for E-S0 galaxies:   HCGs (open circles), KPGs (open triangles),  
and  KIGs (plus signs). \label{ea4twist}}
\end{figure*}

In Figure~\ref{ea4twist}, we show how the isophote parameter $a_{4}$
and twist $\theta$ vary with the ellipticity difference
$\Delta\epsilon$=$\epsilon_{max}$-$\epsilon_{min}$. Values of
$\Delta\epsilon>0$ indicate that the galaxies are generally rounder
in their centers than in their periphery. Large values in
$\Delta\epsilon$ together with large $\mid a_{4}\mid >$0.7 and
$\theta >$20$^\circ$ suggest the galaxies were possibly affected by
interactions. No significant differences are observed between the HCGs
and the KPGs, suggesting, once again, similar formation mechanisms.

In Figure~\ref{comp}, we show the distribution of the morphology of
the galaxies in the different catalogs. We observe a clear
increase in the number of earlier-type galaxies among the HCGs compared with the KPGs
and KIGs. The fact that we found a higher number of S0 galaxies among 
the HCGs than among the KPGs suggests interactions and mergers are
possible mechanisms responsible for forming these galaxies. At the
same time, the fact that we also find S0 galaxies among the KPGs
suggests the environments of these galaxies must have some level of
similarity. For example, one may assume they are different
structures forming in a common or comparable low-density
environment: both form at the periphery of large-scale
structures.

\subsubsection{Intermediate-type (Sa$-$Sb) galaxies}

\begin{figure*}
\epsscale{0.7} \plotone{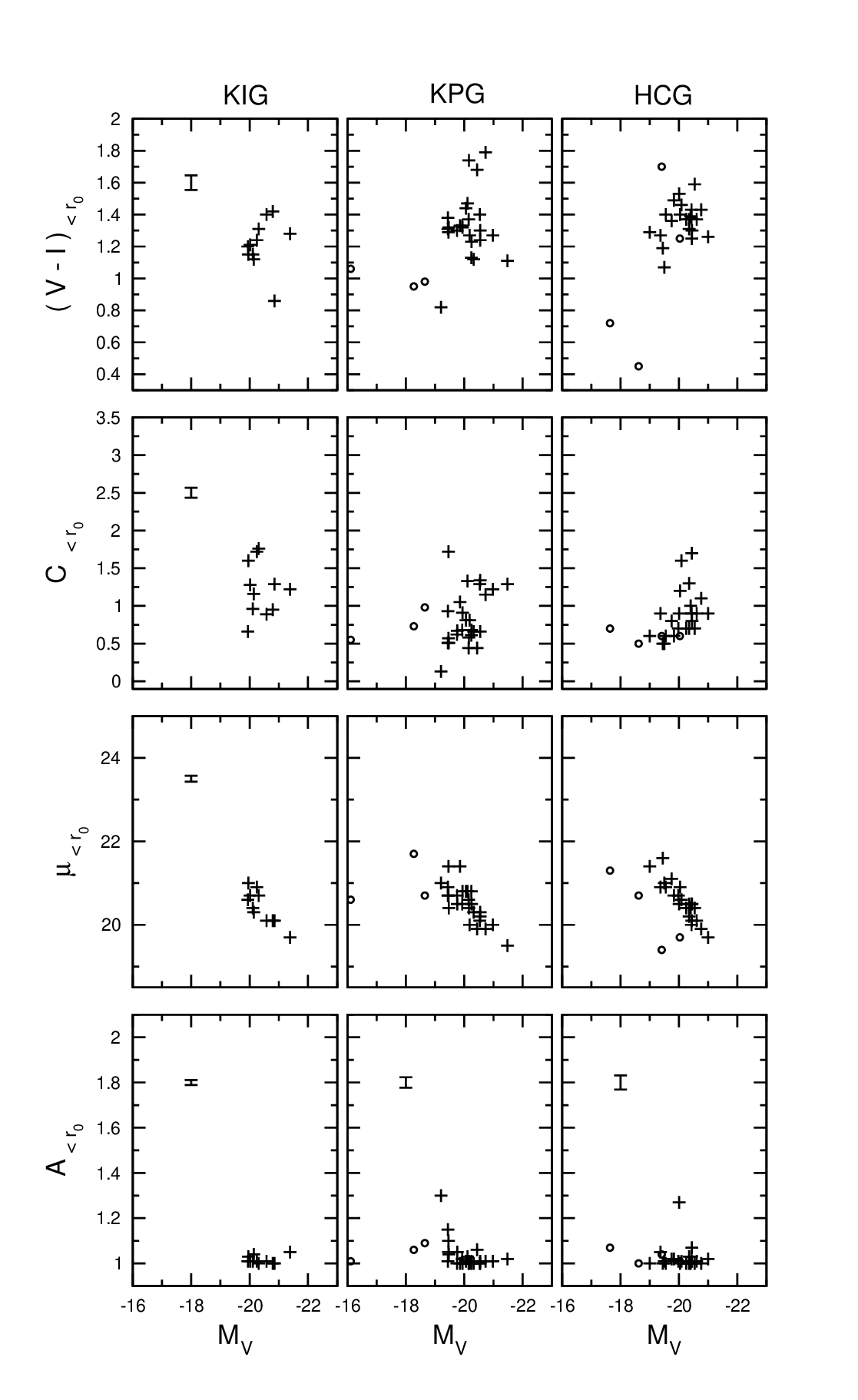} \caption{Variations in
intermediate-type galaxies of isophotal parameters and asymmetry as
a function of absolute magnitude in $V$ inside $r_0$ for $-$from
left to right $-$KIG, KPG, and HCG interdiate-type galaxies. The
parameters are measured  inside the average half-radius $r_0 = 5$
kpc. For smaller size galaxies (open circle), the average half-radius
is $r_0 = 2.5$ kpc. \label{interIn}}
\end{figure*}

\begin{figure*}
\epsscale{0.7} \plotone{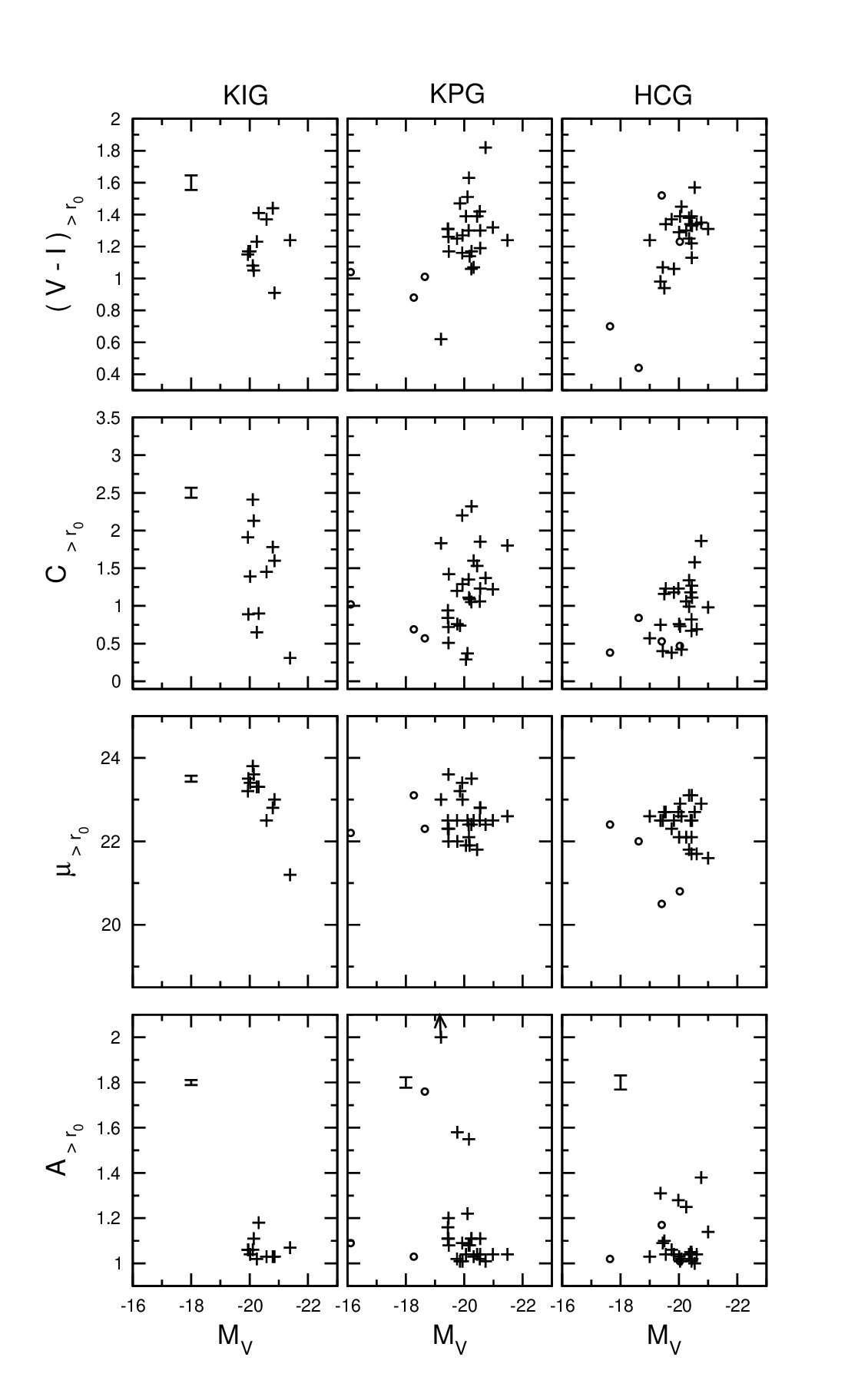} \caption{Variations in
intermediate-type galaxies of isophotal parameters and asymmetry as
a function of absolute magnitude in $V$ inside $r_0$ for $-$from
left to right $-$KIG, KPG, and HCG intermediate-type galaxies. The
parameters are measured  outside the average half-radius $r_0 = 5$
kpc. For smaller size galaxies (open circle), the average half-radius
is $r_0 = 2.5$ kpc. \label{interOut}}
\end{figure*}

In Figures~\ref{interIn} and \ref{interOut}, we show the variations in
the Sa$-$Sb group of the isophotal parameters internal
(Figure~\ref{interIn}) and external (Figure~\ref{interOut}) to
$r_0$. In Figure~\ref{interIn}, the KIGs tend to be slightly
brighter than the KPGs and slightly bluer than the HCGs. 
This is confirmed by statistical tests
(Table~\ref{tabl10}). However, the difference in luminosity may be
due to the fact that there are no small-size Sa$-$Sb galaxies in the
KIGs as compared to the KPGs and HCGs (clearly visible in
Figure~\ref{interIn}). Indeed, when we compare the magnitudes inside
$R_e$ the differences vanish (Table~\ref{tabl11}).

We see a trend for the HCGs to be redder than the KPGs or KIGs 
(Table~\ref{tabl10}). This suggests slightly older nuclear
stellar populations in the HCG galaxies. However, statistical
tests are inconclusive on this matter, except between the HCGs and
KIGs inside $r_0$. The trend toward redder color for the HCG is also
visible using $R_e$, but again statistical tests are
inconclusive (Table~\ref{tabl11}).

Also from Figure~\ref{interIn} and Table~\ref{tabl10} one can see
that the KIG galaxies are more concentrated than the HCGs and KPGs
inside $r_0$. In terms of $R_{90\%}/R_{50\%}$, the statistical tests
only support a difference in concentration between the KIG and KPG
galaxies (Table~\ref{tabl11}). However, the HCG galaxies are
observed to have a greater $R_e$ than the KIG galaxies
(Table~\ref{tabl11}) and to have lower surface brightness at $R_e$
(Table~\ref{tabl11}).

In Figure~\ref{interOut}, the trend in concentration seems to
continue outside $r_0$: the HCG galaxies seem less concentrated than
the galaxies in the other two samples (Table~\ref{tabl10}). Also in
Table~\ref{tabl10} we find differences in surface brightness
outside the half-radius, the HCGs and KPGs having higher surface
brightness than the KIG galaxies.

The differences observed suggest different distributions in mass. In
particular, the intermediate KIG galaxies seem smaller in size and
more compact than the KPG and HCG galaxies. This may be explained by
the isolation status of the KIG: stars in galaxies that have 
experienced interactions are expected to occupy higher energy orbits
than those in galaxies that formed in isolation, and consequently isolated
galaxies may be expected to be more compact or less spatially
extended.

In Figure~\ref{interIn}, no difference is observed in the asymmetry
level. However, outside the half-radius, Figure~\ref{interOut}, the
KPG galaxies tend to be slightly more asymmetric than the HCG
galaxies, even though this is not confirmed by the statistical test
(Table~\ref{tabl10}).

\subsubsection{Late-type (Sbc$-$Im) galaxies}

\begin{figure*}
\epsscale{0.7} \plotone{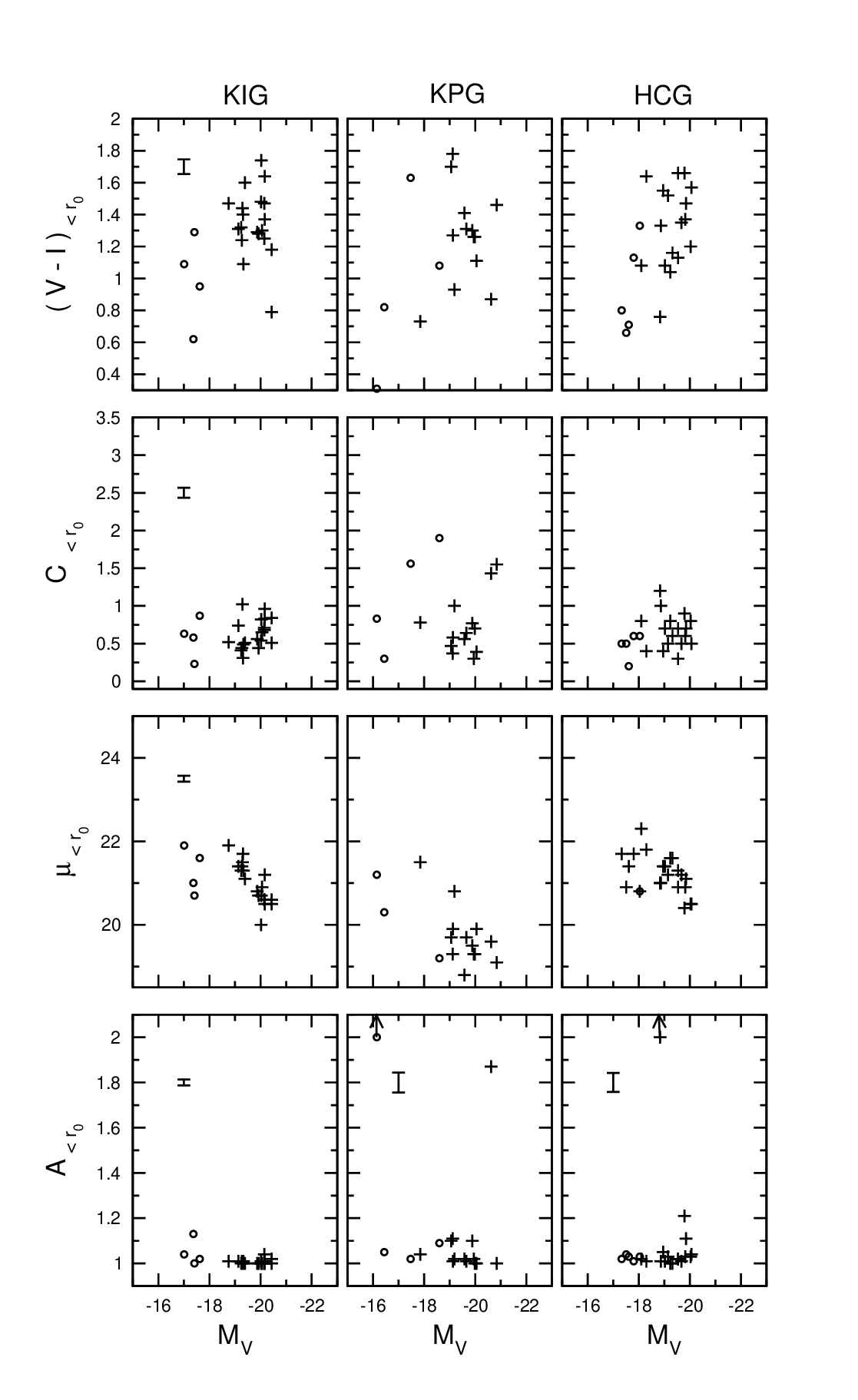} \caption{Variations in late-type
galaxies of isophotal parameters and asymmetry as a function of
absolute magnitude in $V$ inside $r_0$ for $-$from left to right
 $-$KIG, KPG, and HCG late-type galaxies. The parameters are measured
inside the average half-radius $r_0 = 5$ kpc. For smaller size
galaxies (open circle), the average half-radius is $r_0 = 2.5$ kpc.
\label{lateIn}}
\end{figure*}

\begin{figure*}
\epsscale{0.7} \plotone{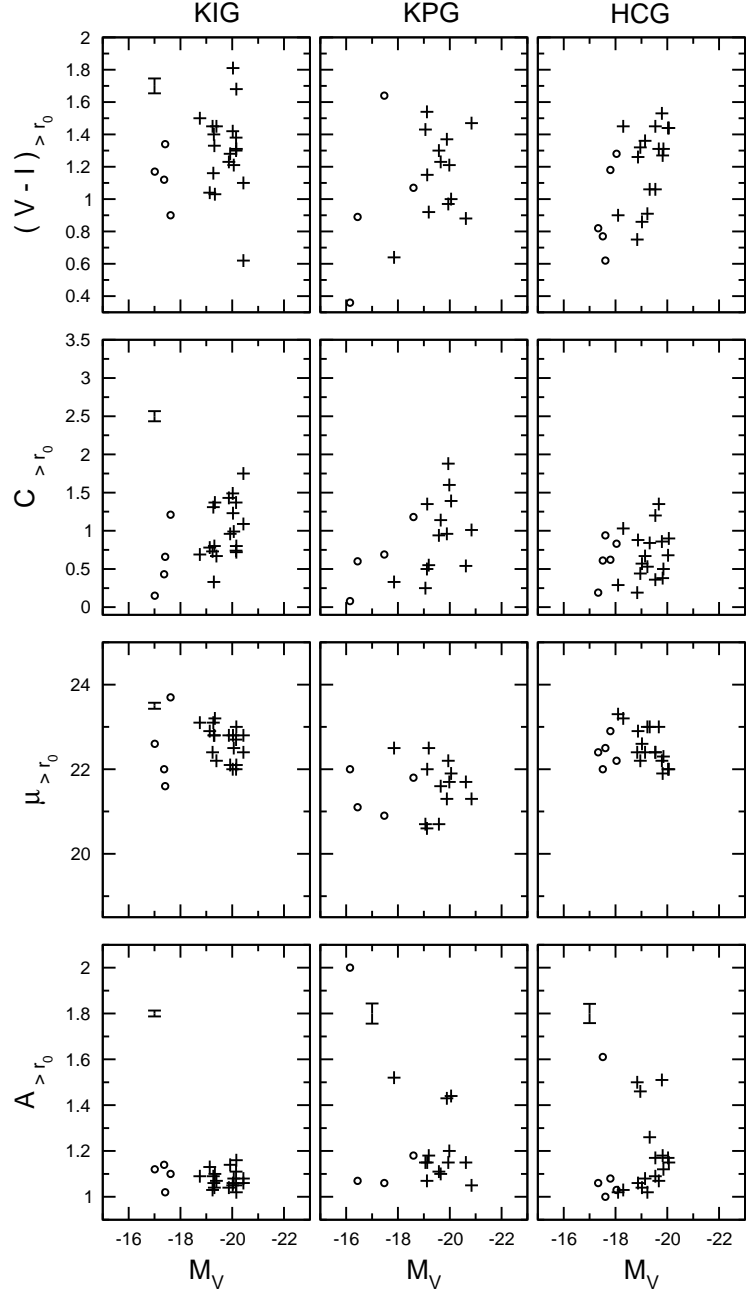} \caption{Variations in late-type
galaxies of isophotal parameters and asymmetry as a function of
absolute magnitude in $V$ inside $r_0$ for $-$from left to right 
$-$KIG, KPG, and HCG late-type galaxies. The parameters are measured
outside the average half-radius $r_0 = 5$ kpc. For smaller size
galaxies (open circle), the average half-radius is $r_0 = 2.5$ kpc.
\label{lateOut}}
\end{figure*}

In Figures~\ref{lateIn} and \ref{lateOut}, we show the variations for
the late-type galaxies of the isophotal parameters internal
(Figure~\ref{lateIn}) and external (Figure~\ref{lateOut}) to $r_0$.
In this group, we observe no obvious differences between the
different parameters. The statistical tests (see Table~\ref{tabl12})
suggest small differences between the HCG and KIG galaxies in terms
of magnitudes, with the KIG galaxies being slightly brighter than
the HCG galaxies both inside $r_0$ and $R_e$.

The HCG galaxies also seem to have lower surface brightness inside
$r_0$ than the KPG galaxies and to be less concentrated than the KIG
galaxies outside $r_0$. The KPG galaxies seem to be bluer than the
HCG inside $R_e$ and to be smaller than the KIG galaxies. We do not
find any other differences based on $R_e$ among the samples (see
Table~\ref{tabl13}). In terms of size and concentration, the trends
seem to go contrary to what is observed for the intermediate types.

The most significant differences observed are in the level of
asymmetry: the KIG galaxies turned out to be more symmetric than the
KPG or the HCG galaxies. The asymmetry level does not seem
significatively different between the HCG and KPG galaxies.

\subsection{Origin of the asymmetries in galaxies}

So far, our analysis has shown differences in the characteristics
of the galaxies that are consistent with evidence for interaction
effects due to their different environments. However, the fact that
we observe different behaviors between morphology groups suggests we
must be careful in our interpretation of asymmetries in terms of
interactions. For example, in intermediate- and late-type spiral
galaxy asymmetric features may be related to internal processes,
like density waves or stochastic star formation propagation, which
are not necessarily produced by interactions. Moreover, in multiple
systems such as in compact groups (or clusters of galaxies),  
a sequence of interaction events may exist that are correlated with the
morphology of the galaxies: early-type galaxies, for example,  may
have entered the systems before late-type ones and would be expected
to show less evidence of interactions than spirals for this reason.

\begin{figure*}
\epsscale{1.0}
\plotone{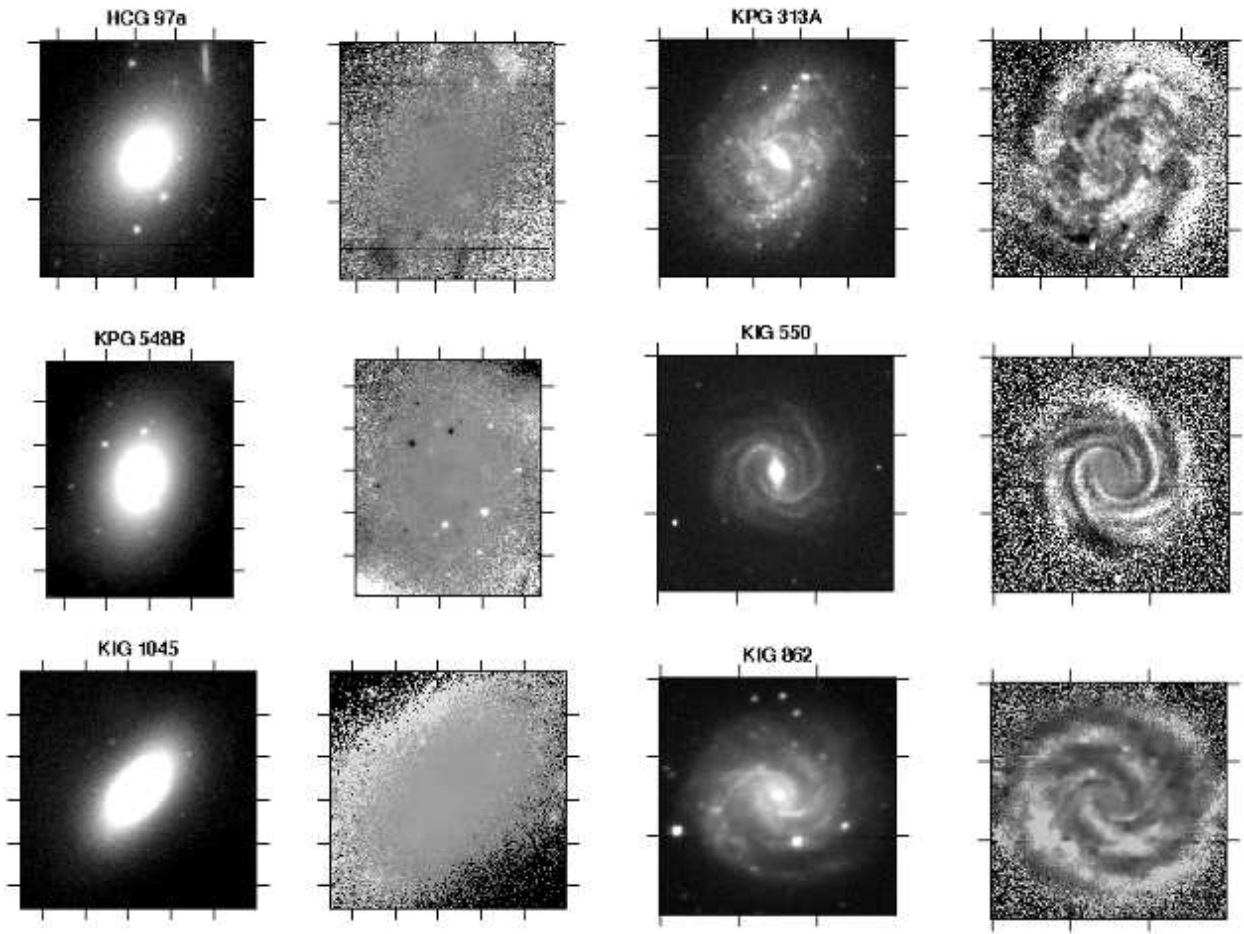}
\caption{Examples of type 1 asymmetries Left: symmetric galaxies; right:  
galaxies where the asymmetries are intrinsic, related to star
formation regions and/or spiral arm structures. The $V$ images are displayed in
logarithmic scales together with their residual images.
\label{type1}}
\end{figure*}

\begin{figure*}
\epsscale{1.0}
\plotone{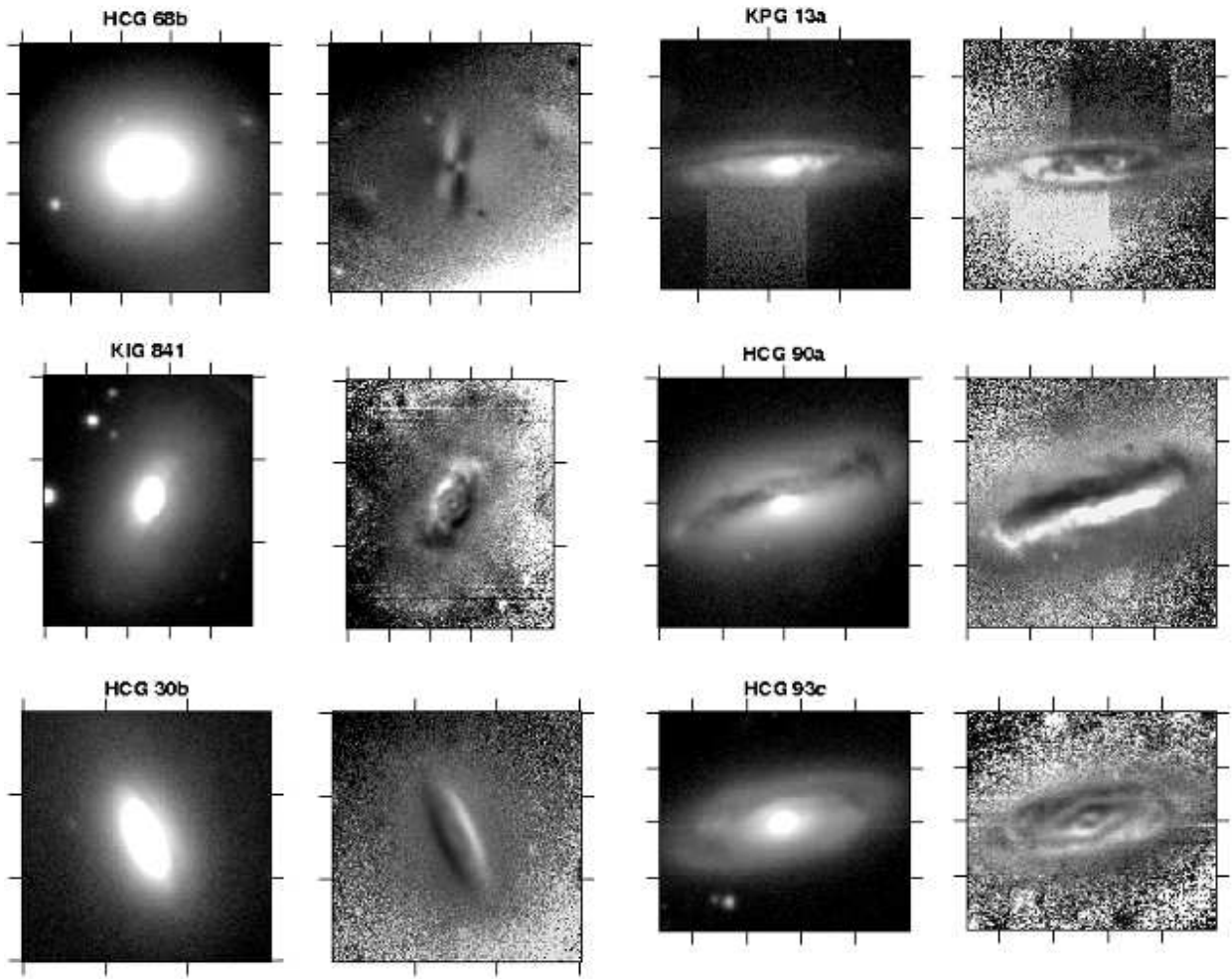}
\caption{Examples of type 2 asymmetries: asymmetry is related to dust lanes in the
disk and/or due to the inclination of the disk on the sky. The images are displayed
in the same way as in Figure~\ref{type1}.
\label{type2}}
\end{figure*}

\begin{figure*}
\epsscale{1.0}
\plotone{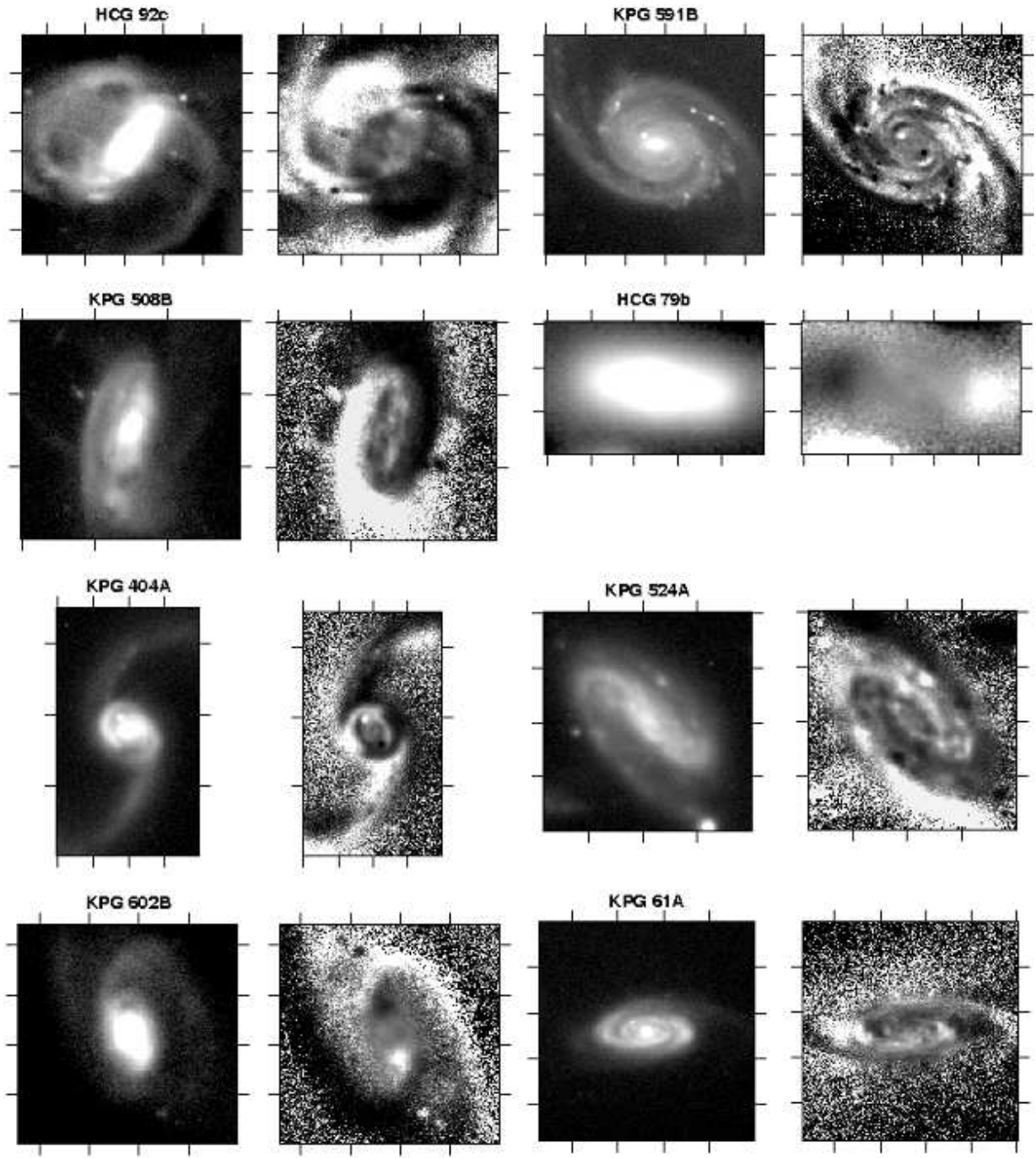}
\caption{Examples of type 3 asymmetries. These are obvious cases of asymmetries related
to galaxies interactions. The images are displayed in the same way as in Figure~\ref{type1}.
\label{type3}}
\end{figure*}

\begin{figure*}
\epsscale{1.0}
\plotone{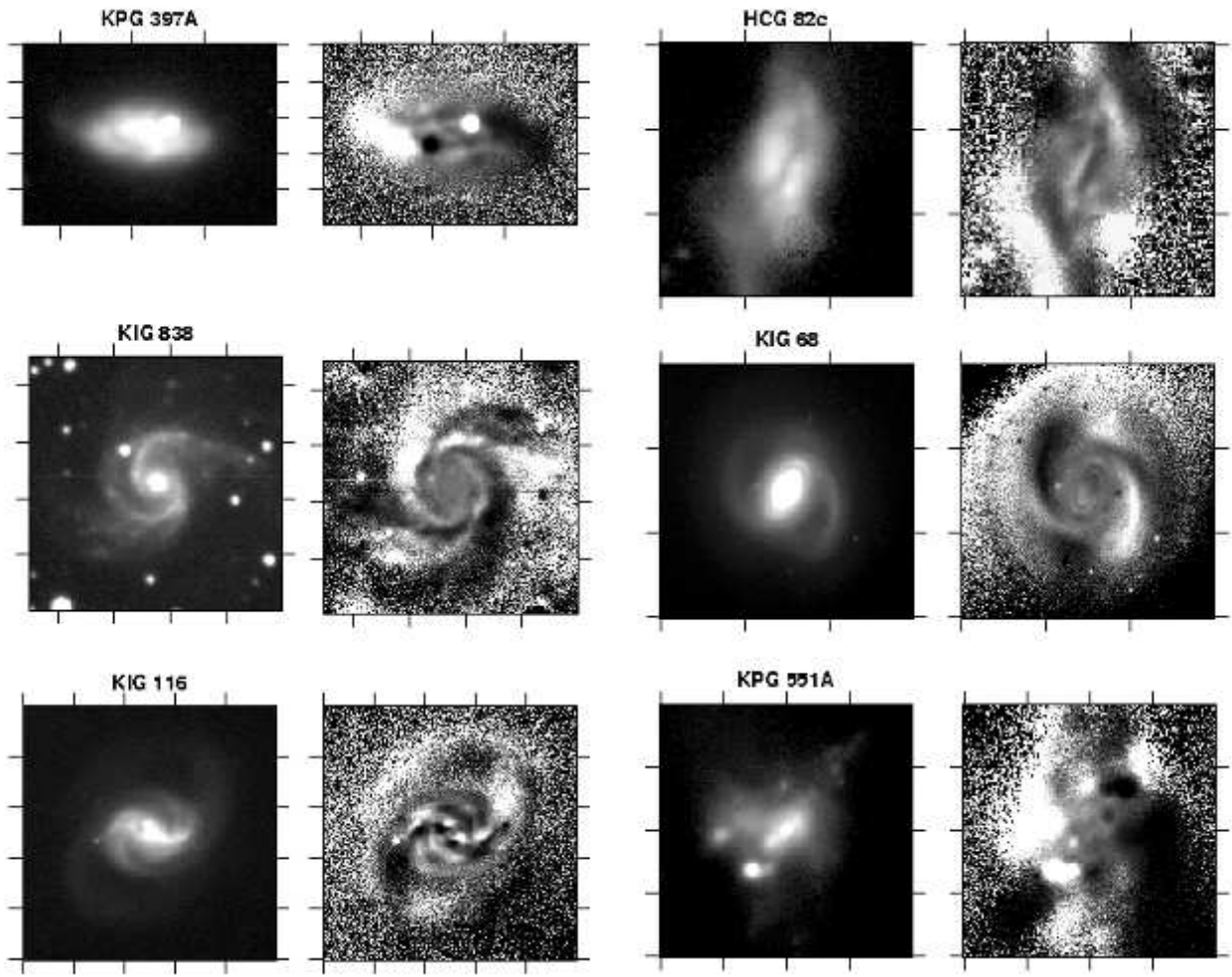}
\caption{Examples of type 4 asymmetries. Asymmetric structures appear, but their cause
is not obvious. The images are displayed in the same way as in Figure~\ref{type1}.
\label{type4}}
\end{figure*}

\begin{figure*}
\epsscale{0.8}
\plotone{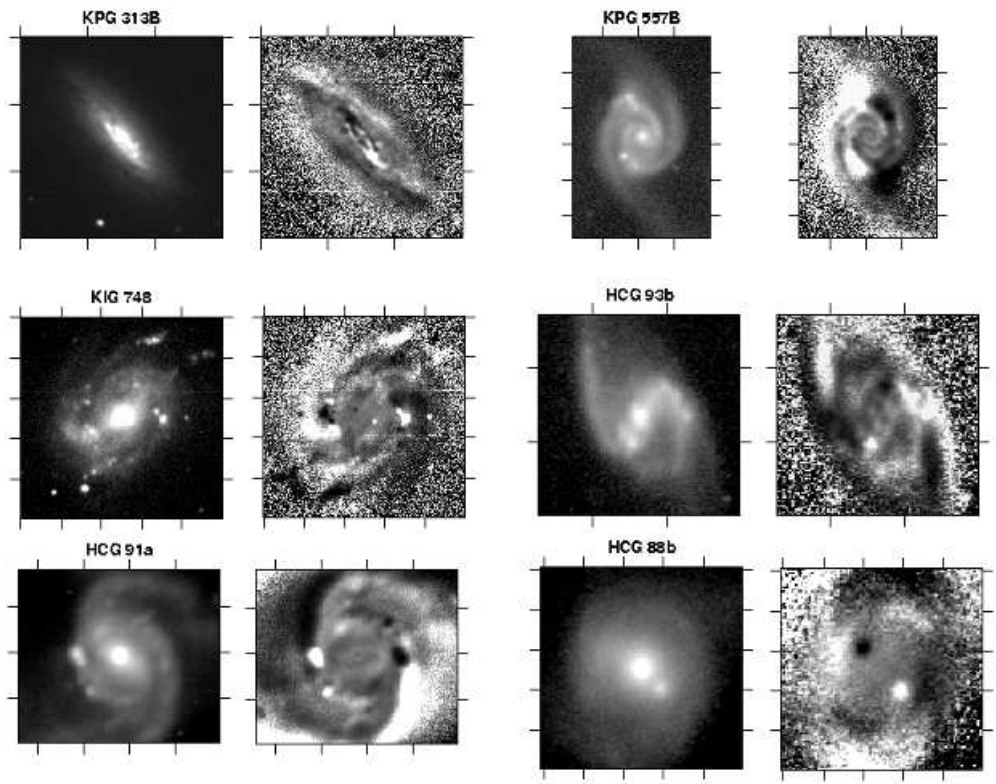}
\caption{Examples of type 5 asymmetries. A companion galaxy appears near the center.
The images are displayed in the same way as in Figure~\ref{type1}.
\label{type5}}
\end{figure*}

\begin{figure*}
\epsscale{1.0}
\plotone{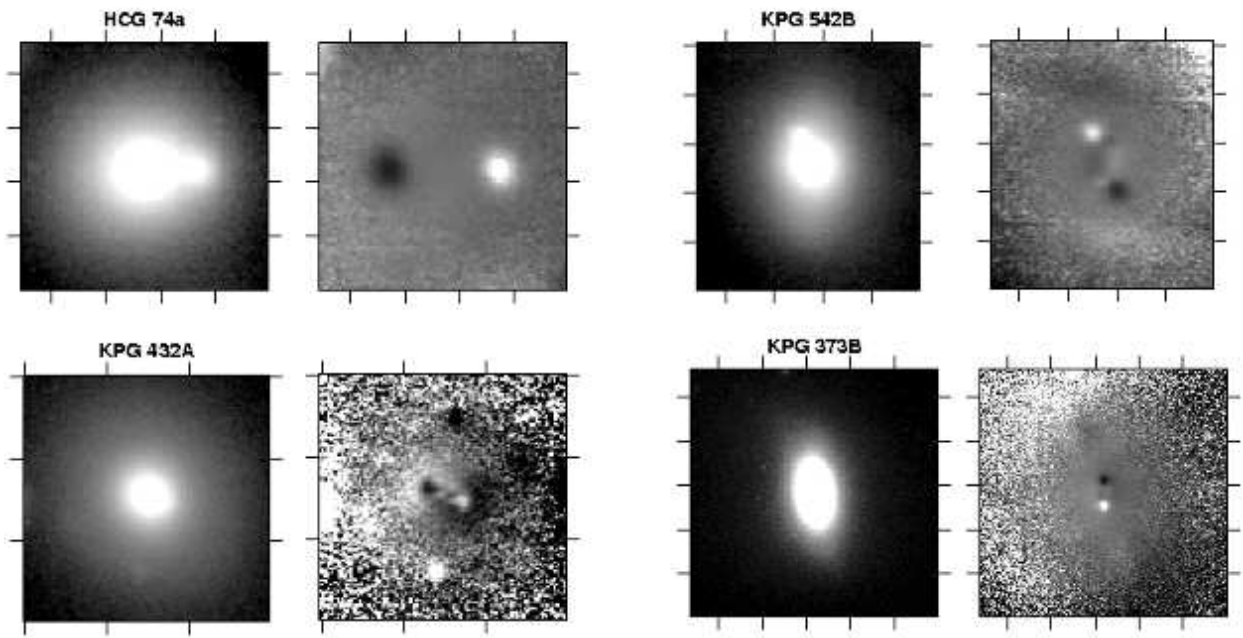}
\caption{ Examples of
type 6 asymmetries. A possible double nucleus is observed in these images. The
images are displayed in the same way as in Figure~\ref{type1}.
\label{type6}}
\end{figure*}

In order to better determine the origin of the asymmetries observed
in the various galaxies of our sample, we have 
meticulously reinspected the residual images produced by our asymmetry analysis
and redistributed the galaxies in our sample in six different types of
asymmetry, independent of the morphology. In type~1, we have put all
the ``symmetric'' galaxies or galaxies with ``intrinsic''
asymmetries related to star formation clumps and/or spiral arms.
Examples of galaxies with a type~1 asymmetry are shown in
Figure~\ref{type1}. In type~2, we have regrouped all the galaxies
where the asymmetry is possibly due to dust or to the inclination of
the galaxy on the plane of the sky. Examples of galaxies with a
type~2 asymmetry are presented in Figure~\ref{type2}. In type~3, we
find the most obvious evidence of galaxy interactions under the
forms of tidal tails, plumes, connecting bridges or a common envelop
between galaxies. Examples of galaxies with a type~3 asymmetry are
presented in Figure~\ref{type3}. We put galaxies that that are highly 
asymmetric, but for which the cause is not obvious in type~4.
Examples of galaxies of this type can be found in
Figure~\ref{type4}. In type~5, we have regrouped the cases where the
asymmetry may be due to a smaller mass satellite galaxy. Examples of
galaxies showing a type~5 asymmetry are shown in Figure~\ref{type5}.
Finally, in type~6 we have regrouped the cases where the asymmetry
is accompanied by a possible double nucleus. Examples of galaxies
with this last type of asymmetry are shown in Figure~\ref{type6}.

\begin{figure*}
\epsscale{0.9}
\plotone{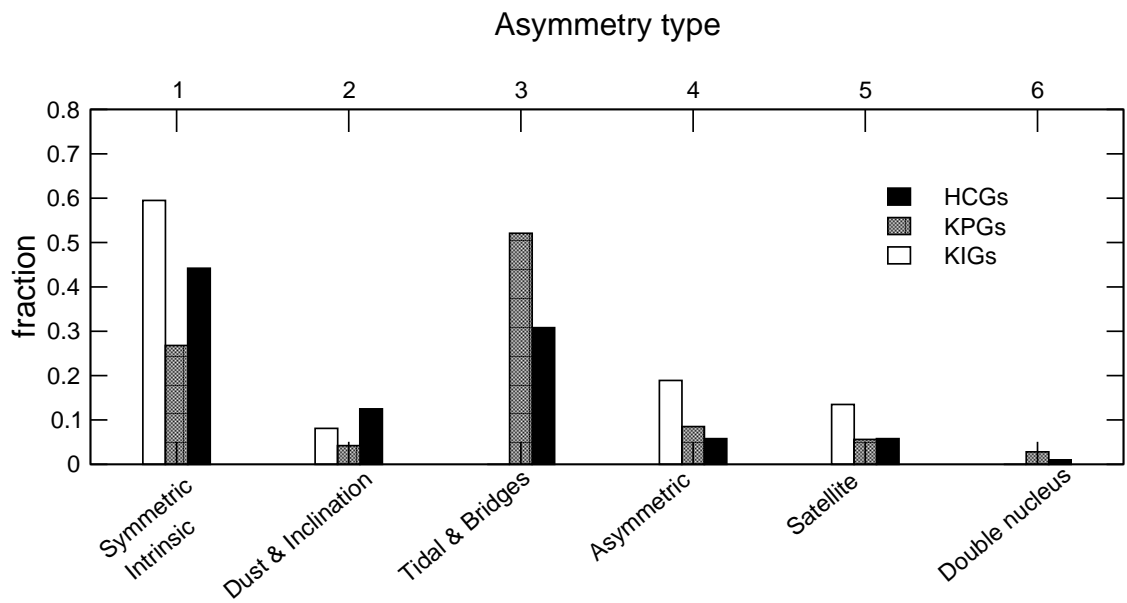}
\caption{ Distribution of the
different asymmetry types in different environments.
\label{distAsy}}
\end{figure*}

The distribution of asymmetry types in the different samples is
presented in Figure~\ref{distAsy}. In the KIG sample, 60\% of the
galaxies have an asymmetry of type~1 and 8\% show an asymmetry of
type~2. Therefore, slightly less than 70\% of the KIG galaxies are
unperturbed. In this group, we do find some ``asymmetric'' galaxies;
however they are either of type~4 (19\%) or of type~5 (13\%). In
general, and as expected, evidence of interactions is largely
missing in the KIGs.

The contrast with the KPGs is significant; as much as 52\% are
classified as type~3, which are obvious cases of recent
interactions. Of the remaining asymmetric galaxies, 8\% are
classified as type~4, 6\% as type~5 and another 3\% as type~6. The
rest of the galaxies are either type~1 (27\%) or type~2 (4\%).
Therefore, almost 70\% of the KPG galaxies  show asymmetries
consistent with ``genuine'' interactions.

In the case of the HCGs, 31\% are classified as type~3, 6\% as type~4,
6\% as type~5, and 1\% as type~6, summing up the evidence for
genuine interactions to 44\%. The number of ``symmetric'' galaxies,
44\% of type~1 and 12\% of type~2, is consequently higher than that in
the KPGs.

\section{Discussion and conclusion}

Through our analysis we have found that galaxies in close pairs show
more frequent signs of interactions at a higher level than those in
compact groups. This may seem somewhat contradictory. If interaction
between galaxies is favored in high-density environments with low
velocity dispersion, should we not expect evidence for such
processes to be more obvious in multiple systems like CGs? A
possible answer to this apparent contradiction can be found in our
isophotal analysis. Indeed, we have seen that the HCG galaxies tend
to be redder in their central part and less compact in their
periphery than the KPG galaxies, which is consistent with older
central stellar populations and dynamically more relaxed orbits as a
whole in the HCG galaxies than in the KPG galaxies. These
observations, together with the presence of asymmetries at a lower
level  in the HCG galaxies, suggest CGs are found in a more advanced
stage of interaction than pairs of galaxies. One possible
explanation is that these structures formed at different epochs: CGs
would have formed in the recent past, while close pairs would have
formed even more recently.

The alternative interpretation is to assume that the evolution of
galaxies is accelerated in CGs: the galaxies in CGs formed at the
same time as those in close pairs, but they evolved faster due to
multiple interactions. However, based on our observations, such an 
alternative seems less probable. In particular, we observe similar
properties for the E$-$S0 in the HCGs and KPGs which suggest similar
formation mechanisms. The higher number of such galaxies in the HCGs
(see Figure~\ref{comp}), therefore, can only be the result of
originally higher matter density: in denser regions, a high number of
galaxies are formed, which can eventually interact to build larger and more
complex structures like CGs, while in less dense environments, a few
galaxies are formed and it can take longer for these galaxies to
interact with neighbors.

On the other hand, the fact that many S0 galaxies can also be found among the
KPGs suggests their environment must have some level of similarity
with that of the HCGs. The common property is that both systems are
examples of structures forming in relatively low density
environments; that is, both form relatively late at the
periphery of large-scale structures.

The cosmological model that better fits our observations is one
where the formation of structures is a biased process. As a
consequence, it is expected that massive structures, which formed in
originally denser regions, must assemble their components at earlier
epochs than less massive ones. If we also assume the formation
process of structures to be continuous in time, then we must now
expect to observe smaller mass structures like CGs and pairs of
galaxies to form at the periphery of the larger-scale structures.

\section{ACKNOWLEDGMENTS}

\acknowledgments We thank the CATT of San Pedro M\'artir for the
observing time given on the 1.5 m telescope to realize this project
and all the personel of the observatory for their support. We also
thank an anonymous referee for important comments and suggestions.
This research has made use of SAOImage DS9, developed by the Smithsonian
Astrophysical Observatory and FTOOLS
(http://heasarc.gsfc.nasa.gov/ftools/), \citep{blackburn95}, and TOPCAT
software provided by the UK's AstroGrid Virtual Observatory Project,
which is funded by the Science and Technology Facilities Council and
through the EU's Framework 6 programs.

\clearpage


\begin{deluxetable}{lrrrlrrrrrr}

\tabletypesize{\scriptsize}
\tablecaption{Properties of the observed KIG galaxies}
\tablewidth{0pt}
\tablehead{
\colhead{Name} & \colhead{ R.A.}      &     \colhead{ Decl.}      & \colhead{$v_{vir}$}         &  \colhead{Morph.}    &    \colhead{$C$}     &  \colhead{$C$}    &   \colhead{$V-I$}   &  \colhead{$V-I$}  &    \colhead{$A$}   &  \colhead{$A$}  \\
\colhead{    } &  \colhead{(J$2000$)} &     \colhead{(J$2000$)}  & \colhead{(km\ s$^{-1}$)}    &  \colhead{Type}      &  \colhead{$<r_0$}    &  \colhead{$>r_0$} &  \colhead{$<r_0$}   &  \colhead{$>r_0$} &  \colhead{$<r_0$}  &  \colhead{$>r_0$}    \\
\colhead{(1)}  & \colhead{(2)}        & \colhead{(3)}            & \colhead{(4)}               & \colhead{(5)}        & \colhead{(6)}        & \colhead{(7)}     & \colhead{(8)}       & \colhead{(9)}     & \colhead{(10)}     & \colhead{(11)}    \\
}
\startdata

KIG~ 33 &   $    0\phn43\phn28  $   &   $   -00\phn07\phn30 $   &   4134    &   SBb*    &   0.96    &   2.41    &   1.15    &   1.08    &   1.01    &   1.06    \\
KIG~ 53 &   $    1\phn30\phn47  $   &   $   21\phn26\phn26  $   &   3199    &   SBbc    &   0.49    &   1.37    &   1.09    &   1.03    &   1.01    &   1.10    \\
KIG~ 56 &   $    1\phn36\phn00  $   &   $   00\phn39\phn49  $   &   5114    &   SBb     &   0.66    &   1.91    &   1.20    &   1.15    &   1.01    &   1.06    \\
KIG~ 61 &   $    1\phn42\phn27  $   &   $   26\phn08\phn35  $   &   3986    &   SBab    &   1.76    &   0.90    &   1.31    &   1.41    &   1.00    &   1.18    \\
KIG~ 68 &   $    1\phn53\phn13  $   &   $   04\phn11\phn45  $   &   1686    &   SBa     &   1.60    &   0.89    &   1.15    &   1.17    &   1.03    &   1.06    \\
KIG~ 116    &   $   23\phn58\phn32  $   &   $   26\phn12\phn51  $   &   3363    &   SBab    &   1.16    &   2.13    &   1.12    &   1.05    &   1.04    &   1.11    \\
KIG~ 123    &   $    3\phn12\phn51  $   &   $   04\phn42\phn25  $   &   5853    &   SBbc    &   0.68    &   0.74    &   1.25    &   1.38    &   1.00    &   1.05    \\
KIG~ 467    &   $   11\phn09\phn16  $   &   $   36\phn01\phn16  $   &   6560    &   SB0     &   0.97    &   1.65    &   1.36    &   1.38    &   1.01    &   1.03    \\
\enddata
\tablecomments{Columns: (1) catalog galaxy identification;
(2) right ascension from HyperLeda ($0^{h}00^{m}00^{s}$);
(3) declination from HyperLeda ($0^{\circ}00'00''$);
(4) radial velocity corrected for infall Local Group toward Virgo from HyperLeda;
(5) morphological type as determined in this work;
(6) concentration inside $r_0$;
(7) concentration outside $r_0$;
(8) average $V-I$ color inside $r_0$;
(9) average $V-I$ color outside $r_0$;
(10) average asymmetry inside $r_0$; and
(11) average asymmetry outside $r_0$.
 }
\label{tabl1}
\end{deluxetable}

\begin{deluxetable}{lrrrlrrrrrr}

\tabletypesize{\scriptsize}
\tablecaption{Properties of the observed KPG galaxies}
\tablewidth{0pt}
\tablehead{
\colhead{Name} & \colhead{ R.A.}      &     \colhead{ Decl.}      & \colhead{$v_{vir}$}         &  \colhead{Morph.}    &    \colhead{$C$}     &  \colhead{$C$}    &   \colhead{$V-I$}   &  \colhead{$V-I$}  &    \colhead{$A$}   &  \colhead{$A$}  \\
\colhead{    } &  \colhead{(J$2000$)} &     \colhead{(J$2000$)}  & \colhead{(km\ s$^{-1}$)}    &  \colhead{Type}      &  \colhead{$<r_0$}    &  \colhead{$>r_0$} &  \colhead{$<r_0$}   &  \colhead{$>r_0$} &  \colhead{$<r_0$}  &  \colhead{$>r_0$}    \\
\colhead{(1)}  & \colhead{(2)}        & \colhead{(3)}            & \colhead{(4)}               & \colhead{(5)}        & \colhead{(6)}        & \colhead{(7)}     & \colhead{(8)}       & \colhead{(9)}     & \colhead{(10)}     & \colhead{(11)}    \\
}
\startdata

KPG~ 13A    &   $    0\phn36\phn52  $   &   $   23\phn59\phn27  $   &   4701    &   Sb  &   0.93    &   0.94    &   1.38    &   1.31    &   1.15    &   1.11    \\
KPG~ 13B    &   $    0\phn36\phn52  $   &   $   23\phn59\phn05  $   &   4573    &   S0  &   0.93    &   0.52    &   0.70    &   0.75    &   1.08    &   1.17    \\
KPG~ 61A    &   $    2\phn16\phn13  $   &   $   32\phn38\phn58  $   &   4916    &   Sb  &   0.68    &   2.20    &   1.33    &   1.16    &   1.02    &   1.09    \\
KPG~ 61B    &   $    2\phn16\phn21  $   &   $   32\phn39\phn59  $   &   4752    &   SB0 &   1.27    &   1.02    &   1.22    &   1.18    &   1.01    &   1.01    \\
KPG~ 75A    &   $    2\phn45\phn10  $   &   $   32\phn59\phn23  $   &   5099    &   SBa &   1.34    &   1.23    &   1.30    &   1.30    &   1.01    &   1.11    \\
KPG~ 75B    &   $    2\phn45\phn14  $   &   $   32\phn58\phn41  $   &   5169    &   SBb &   0.57    &   0.72    &   1.29    &   1.26    &   1.05    &   1.20    \\
KPG~ 99A    &   $    4\phn30\phn40  $   &   $   00\phn39\phn43  $   &   3590    &   E   &   1.47    &   0.59    &   1.28    &   1.26    &   1.03    &   1.18    \\
KPG~ 99B    &   $    4\phn30\phn44  $   &   $   00\phn39\phn53  $   &   3411    &   E   &   1.50    &   0.63    &   1.24    &   1.20    &   1.04    &   1.13    \\
\enddata
\tablecomments{Columns: (1) catalog galaxy identification;
(2) right ascension from HyperLeda ($0^{h}00^{m}00^{s}$);
(3) declination from HyperLeda ($0^{\circ}00'00''$);
(4) radial velocity corrected for infall Local Group toward Virgo from HyperLeda;
(5) morphological type as determined in this work;
(6) concentration inside $r_0$;
(7) concentration outside $r_0$;
(8) average $V-I$ color inside $r_0$;
(9) average $V-I$ color outside $r_0$;
(10) average asymmetry inside $r_0$; and
(11) average asymmetry outside $r_0$.
 }
\label{tabl2}
\end{deluxetable}

\begin{deluxetable}{lrrrlrrrrrr}

\tabletypesize{\scriptsize}
\tablecaption{Properties of the observed HCG galaxies}
\tablewidth{0pt}
\tablehead{
\colhead{Name} & \colhead{ R.A.}      &     \colhead{ Decl.}      & \colhead{$v_{vir}$}         &  \colhead{Morph.}    &    \colhead{$C$}     &  \colhead{$C$}    &   \colhead{$V-I$}   &  \colhead{$V-I$}  &    \colhead{$A$}   &  \colhead{$A$}  \\
\colhead{    } &  \colhead{(J$2000$)} &     \colhead{(J$2000$)}  & \colhead{(km\ s$^{-1}$)}    &  \colhead{Type}      &  \colhead{$<r_0$}    &  \colhead{$>r_0$} &  \colhead{$<r_0$}   &  \colhead{$>r_0$} &  \colhead{$<r_0$}  &  \colhead{$>r_0$}    \\
\colhead{(1)}  & \colhead{(2)}        & \colhead{(3)}            & \colhead{(4)}               & \colhead{(5)}        & \colhead{(6)}        & \colhead{(7)}     & \colhead{(8)}       & \colhead{(9)}     & \colhead{(10)}     & \colhead{(11)}    \\
}
\startdata

HCG~ 01a    &   $    0\phn26\phn07  $   &   $   25\phn43\phn31  $   &   10332   &   Sc  &   0.63    &   0.38    &   1.37    &   1.27    &   1.03    &   1.18    \\
HCG~ 01b    &   $    0\phn26\phn06  $   &   $   25\phn43\phn10  $   &   10361   &   E   &   0.94    &   0.16    &   1.28    &   1.19    &   1.13    &   1.25    \\
HCG~ 01c    &   $    0\phn25\phn54  $   &   $   25\phn43\phn24  $   &   10151   &   E   &   1.19    &   0.77    &   1.35    &   1.33    &   1.00    &   1.00    \\
HCG~ 10a    &   $    1\phn26\phn22  $   &   $   34\phn42\phn11  $   &   5269    &   SBb &   1.07    &   1.86    &   1.43    &   1.35    &   1.00    &   1.38    \\
HCG~ 10c    &   $    1\phn26\phn19  $   &   $   34\phn45\phn15  $   &   4781    &   SB0 &   1.26    &   0.89    &   1.41    &   1.36    &   1.01    &   1.02    \\
HCG~ 10d    &   $    1\phn26\phn31  $   &   $   34\phn40\phn32  $   &   4741    &   Scd &   1.03    &   0.88    &   1.33    &   1.26    &   1.01    &   1.06    \\
HCG~ 14a    &   $    1\phn59\phn52  $   &   $   -07\phn05\phn12 $   &   5261    &   SBb &   0.58    &   1.23    &   1.40    &   1.34    &   1.00    &   1.04    \\
HCG~ 14b    &   $    1\phn59\phn50  $   &   $   -07\phn03\phn33 $   &   5825    &   S0  &   0.97    &   0.77    &   1.23    &   1.16    &   1.00    &   1.03    \\
\enddata
\tablecomments{Columns: (1) catalog Galaxy identification;
(2) right ascension from HyperLeda ($0^{h}00^{m}00^{s}$);
(3) declination from HyperLeda ($0^{\circ}00'00''$);
(4) radial velocity from Hickson et al. (1992) and corrected for infall Local Group toward Virgo. ;
(5) morphological type as determined in this work;
(6) concentration inside $r_0$;
(7) concentration outside $r_0$;
(8) average $V-I$ color inside $r_0$;
(9) average $V-I$ color outside $r_0$;
(10) average asymmetry inside $r_0$; and
(11) average asymmetry outside $r_0$.
}
\label{tabl3}
\end{deluxetable}

\begin{deluxetable}{lcccccc}
\tabletypesize{\scriptsize} \tablecaption{Properties of Observed Versus. Catalog Galaxies}
\tablewidth{0pt}
\tablehead{
\colhead{Sample}  & \colhead{$M_{B}$}     & \colhead{$P_{MW}$}     &\colhead{$D_{B}$}     &  \colhead{$P_{MW}$}    &\colhead{$V_{vir}$}     &  \colhead{$P_{MW}$}     \\
\colhead{}        & \colhead{(mag)}       & \colhead{}             &\colhead{(kpc)}       &  \colhead{}            &\colhead{(km s$^{-1}$)} &  \colhead{}        \\
\colhead{(1)}     & \colhead{(2)}         & \colhead{(3)}          &\colhead{(4)}         & \colhead{(5)}          & \colhead{(6)}          &  \colhead{(7)}
}
\startdata
KIG  &  -20.46/-20.30   &  0.070  &  25/22 &  0.153  &  4934/6296 & \underline{0.0003}      \\
KPG  &  -20.35/-20.32   &  0.619  &  23/23 &  0.752  &  5123/6326 &  0.0177               \\
HCG  &  -20.19/-20.01   &  0.119  &  26/24 &  0.050  &  7150/7970 &  0.6847               \\
\enddata
\tablecomments{Columns: (1) sample identification; (2)
median $M_{B}$ value  of the observed/catalog galaxies; (3)
probability $P$ for $M_{B}$; (4) median value  $D_{B}$ of the
observed/catalog galaxies; (5) probability $P$ for $D_{B}$; (6)
median $v_{vir}$ value of the observed/catalog galaxies; and (7)
probability $P$ for $v_{vir}$. $P$ values were obtained from Mann$-$Whitney tests.
Underlined values indicate statistically significant differences between two samples.}
\label{tabl4}
\end{deluxetable}

\begin{deluxetable}{cllccrr}
\tabletypesize{\scriptsize}
\tablecaption{Observing runs}
\tablewidth{0pt} \tablehead{
\colhead{Run}    &  \colhead{Date} & \colhead{CCD} & \colhead{Filters} & \colhead{Seeing} & \colhead{$\sigma_{V}$} & \colhead{$\sigma_{I}$} \\
\colhead{} &  \colhead{}   &  \colhead{}   & \colhead{} & \colhead{(FWHM)}        & \colhead{(mag)}      & \colhead{(mag)} \\
\colhead{(1)}    & \colhead{(2)}   & \colhead{(3)}    & \colhead{(4)}     & \colhead{(5)}   & \colhead{(6)}       & \colhead{(7)}}
\startdata
1 &   2005 Oct            &  Site1     & $V,I$ &  2.4$''$   &  $\pm0.02$ & $\pm0.03$ \\
2 &   2006 Aug            &  Site1     & $V,I$ &  1.6$''$   &  $\pm0.03$ & $\pm0.05$ \\
3 &   2007 Sep            &  Marconi   & $V,I$ &  1.9$''$   &  $\pm0.03$ & $\pm0.04$ \\
4 &   2008 May            &  Site1     & $V,I$ &  1.8$''$   &  $\pm0.11$ & $\pm0.08$ \\
5 &   2008 Jul            &  Site1     & $V,I$ &  1.4$''$   &  $\pm0.05$ & $\pm0.04$ \\
\enddata
\tablecomments{Columns: (1) running number; (2) observation dates; (3) CCD detectors; (4) filters;
(5) average seeing; (6) and (7) calibration uncertainties for $V$ and $I$, respectively }
\label{tabl5}
\end{deluxetable}

\begin{deluxetable}{llrrrrrrrcr}

\tabletypesize{\scriptsize}
\tablecaption{Properties of observed galaxies using effective radius ($1/\eta (R_P)=0.2$)}
\tablewidth{0pt}
\tablehead{
\colhead{Name} & \colhead{ Morph.}  & \colhead{ $R_P$}   & \colhead{$R_P$} &  \colhead{$R_e$}    & \colhead{$R_e$} &  \colhead{$M_eV$ }&  \colhead{$M_eI$} & \colhead{$R_{90}/R_{50}$} &  \colhead{$\mu_V$}            &  \colhead{$V-I$}    \\
\colhead{    } &  \colhead{(type)}  & \colhead{(arcsec)} & \colhead{(kpc)} &  \colhead{(arcsec)} & \colhead{(kpc)} &  \colhead{(mag)}  &  \colhead{(mag)}  & \colhead{}               &  \colhead{(mag arcsec$^{-2}$)}  &  \colhead{(mag)}    \\
\colhead{(1)}  & \colhead{(2)}      & \colhead{(3)}      & \colhead{(4)}   & \colhead{(5)}       & \colhead{(6)}   & \colhead{(7)}     & \colhead{(8)}     & \colhead{(9)}            & \colhead{(10)}                 & \colhead{(11)}
}
\startdata
 & & & & & KIGs & & & & & \\
\hline

  KIG~33 & SBb* & 21.0 &   6 & 11.6 &   3 & -19.57 & -20.59 & 1.7 & 20.5 & 1.02 \\
  KIG~53 & SBbc & 40.3 &   8 & 25.4 &   5 & -19.42 & -20.32 & 1.4 & 21.9 & 0.91 \\
  KIG~56 & SBb & 20.7 &   7 & 9.2 &   3 & -19.61 & -20.60 & 2.2 & 20.9 & 0.99 \\
  KIG~61 & SBab & 10.8 &   3 & 3.3 &   1 & -19.19 & -20.39 & 2.8 & 19.0 & 1.19 \\
  KIG~68 & SBa & 11.2 &   1 & 4.1 & $<$1  & -18.20 & -19.27 & 2.3 & 18.6 & 1.08 \\
  KIG~116 & SBab & 23.2 &   5 & 11.5 &   2 & -19.41 & -20.40 & 1.9 & 20.3 & 0.99 \\
  KIG~123 & SBbc & 70.0 &  26 & 35.9 &  14 & -21.17 & -22.13 & 1.7 & 21.7 & 0.97 \\
\enddata
\tablecomments{Columns: (1) catalog galaxy identification;
(2) morphological type as determined in this work;
(3) petrosian radius defined as $1/\eta (R_P)=0.2$;
(4) petrosian radius in kiloparsec;
(5) effective radius $R_e$;
(6) effective radius in kiloparsec;
(7) absolute magnitude in $V$ inside $R_e$;
(8) absolute magnitude in $I$ inside $R_e$;
(9) concentration index, defined as the ratio of radii that contain 90\% and 50\% of the Petrosian flux;
(10) surface brightness at $R_e$; and
(11) color at $R_e$.
}
\label{tabl6}
\end{deluxetable}

\begin{deluxetable}{llrrrrrrrcr}

\tabletypesize{\scriptsize} \tablecaption{Properties observed
galaxies using effective radius ($1/\eta (R_P)=0.4$)}
\tablewidth{0pt} \tablehead{
\colhead{Name} & \colhead{ Morph.}  & \colhead{ $R_P$}   & \colhead{$R_P$} &  \colhead{$R_e$}    & \colhead{$R_e$} &  \colhead{$M_eV$ }&  \colhead{$M_eI$} & \colhead{$R_{90}/R_{50}$} &  \colhead{$\mu_V$}            &  \colhead{$V-I$}    \\
\colhead{    } &  \colhead{(type)}  & \colhead{(arcsec)} & \colhead{(kpc)} &  \colhead{(arcsec)} & \colhead{(kpc)} &  \colhead{(mag)}  &  \colhead{(mag)}  & \colhead{}               &  \colhead{(mag arcsec$^{-2}$)}  &  \colhead{(mag)}    \\
\colhead{(1)}  & \colhead{(2)}      & \colhead{(3)}      & \colhead{(4)}   & \colhead{(5)}       & \colhead{(6)}   & \colhead{(7)}     & \colhead{(8)}     & \colhead{(9)}            & \colhead{(10)}                 & \colhead{(11)}    \\
}
\startdata
 & & & & & KIGs & & & & & \\
\hline
  KIG~716 & Sc & 4.9 &   3 & 2.0 &   1 & -19.08 & -20.68 & 2.3 & 19.9 & 1.60 \\
  KIG~748 & SBc* & 45.0 &   3 & 26.0 &   2 & -16.64 & -17.54 & 1.6 & 22.2 & 0.90 \\
  KIG~808 & SBc & 26.2 &   3 & 15.5 &   2 & -16.96 & -18.65 & 1.6 & 21.1 & 1.69 \\
  KIG~924 & Sc & 29.3 &   9 & 18.7 &   6 & -19.23 & -20.30 & 1.5 & 22.1 & 1.06 \\
  KIG~931 & SBbc & 15.4 &   5 & 12.9 &   4 & -18.56 & -19.51 & 1.2 & 21.8 & 0.95 \\
  KIG~967 & Sc & 41.4 &   3 & 24.4 &   2 & -16.86 & -17.54 & 1.6 & 21.3 & 0.68 \\
  KIG~983 & Sc & 35.5 &  11 & 21.6 &   7 & -19.93 & -21.31 & 1.5 & 21.6 & 1.38 \\
\enddata
\tablecomments{Columns: (1) catalog galaxy identification;
(2) morphological type as determined in this work;
(3) petrosian radius defined as $1/\eta (R_P)=0.4$;
(4) petrosian radius in kiloparsec;
(5) effective radius, $R_e$;
(6) effective radius in kiloparsec;
(7) absolute magnitude in $V$ inside $R_e$;
(8) absolute magnitude in $I$ inside $R_e$;
(9) concentration index, defined as the ratio of radii that contain 90\% and 50\% of the Petrosian flux;
(10) surface brightness at $R_e$; and
(11) color at $R_e$.
}
\label{tabl7}
\end{deluxetable}

\begin{deluxetable}{lrrr}
\tabletypesize{\scriptsize}
\tablewidth{0pt}
\tablecaption{Properties of Early-type galaxies in different environments}
\tablehead{
\colhead{Property}  &  \colhead{HCGs}   &   \colhead{KPGs} & \colhead{$P_{MW}$}  \\
\colhead{(1)}       &  \colhead{(2)}    &   \colhead{(3)}  & \colhead{(4)}
}
\startdata

$M_{V}$          & -20.3   & -20.4    & 0.4006                   \\
$M_{I}$          & -21.7   & -21.5    & 0.0992                   \\
$\mu_{<r_{0}}$   & 20.4    & 20.4     & 0.2233                   \\
$\mu_{>r_{0}}$   & 22.4    & 23.0     & \underline{0.0026}  \\
$(V-I)_{<r_{0}}$ & 1.34    & 1.25     & \underline{0.0017}  \\
$(V-I)_{>r_{0}}$ & 1.33    & 1.21     & \underline{0.0100}  \\
$C_{<r_{0}}$     & 1.1     & 1.4      & \underline{0.0040}  \\
$C_{>r_{0}}$     & 0.7     & 0.7      & 0.1193                   \\
$A_{<r_{0}}$     & 1.00    & 1.00     & 0.0716                   \\
$A_{>r_{0}}$     & 1.01    & 1.01     & 0.0830                   \\

\enddata
\tablecomments{Columns: (1) properties measured in each sample (all values are medians): absolute magnitude in $V$ inside $r_{0}$,
absolute magnitude in $I$ inside $r_{0}$, surface brightness in $V$ inside and outside $r_{0}$,
$V-I$ color inside and outside $r_{0}$, concentration index inside and outside $r_{0}$,
asymmetry level inside and outside $r_{0}$; (2) and (3) galaxies in HCGs and KPGs, respectively;
(4) $P$ values from Mann$-$Whitney tests, where underlined values indicate statistically significant differences.}
\label{tabl8}
\end{deluxetable}

\begin{deluxetable}{lrrr}
\tabletypesize{\scriptsize}
\tablewidth{0pt}
\tablecaption{Properties of Early-type galaxies as measured using $R_e$}
\tablehead{
\colhead{Property}  &  \colhead{HCGs}   &   \colhead{KPGs} & \colhead{$P_{MW}$}  \\
\colhead{(1)}       &  \colhead{(2)}    &   \colhead{(3)}  & \colhead{(4)}
}
\startdata

$M_V$             & -20.4   & -20.2    & 0.0669              \\
$M_I$             & -21.7   & -21.3    & \underline{0.0368}  \\
$\mu_V$           &  20.2   &  19.5    & \underline{0.0124}  \\
$(V-I)$           &  1.32   &  1.20    & \underline{0.0036}  \\
$R_{90}/R_{50}$   &  2.2    &  2.3     & 0.2320              \\
$R_e$             &  2      &   1      & \underline{0.0002}  \\

\enddata
\tablecomments{ Columns: (1) properties measured in each sample (all values are medians): absolute magnitude in $V$ inside $R_e$,
absolute magnitude in $I$ inside $R_e$, surface brightness in $V$ at $R_e$, $V-I$ color at $R_e$, concentration index, defined by
the ratio of the radii containing 90\% and 50\% of the Petrosian flux, and effective radius in kiloparsec; (2) and (3)
galaxies in HCG and KPG, respectively; (4) $P$ values from Mann$-$Whitney test, where underlined
values indicate statistically significant differences.}
\label{tabl9}
\end{deluxetable}

\begin{deluxetable}{lrrrrrr}
\tabletypesize{\scriptsize}
\tablewidth{0pt}
\tablecaption{Properties of Intermediate-type galaxies in different environments}
\tablehead{
\colhead{Property}  &  \colhead{HCG}  &  \colhead{KPG}  & \colhead{KIG} &  \colhead{HCG$-$KPG} & \colhead{HCG4$-$KIG}  & \colhead{KPG$-$KIG} \\
\colhead{(1)}       & \colhead{(2)}    &  \colhead{(3)}   & \colhead{(4)}  &  \colhead{(5)}     & \colhead{(6)}      & \colhead{(7)}
}
\startdata

$M_{V}$           &  -20.0  &  -20.1  &   -20.3   & 0.3646                   & 0.0631                 & \underline{0.0291}     \\
$M_{I}$           &  -21.4  &  -21.3  &   -21.4   & 0.4051                   & 0.1983                 & 0.1200                 \\
$\mu_{<r_{0}}$    &   20.5  &   20.6  &    20.4   & 0.4315                   & 0.2913                 & 0.2470                 \\
$\mu_{>r_{0}}$    &   22.5  &   22.5  &    23.3   & 0.1697                   & \underline{0.0004}     & \underline{0.0034}     \\
$(V-I)_{<r_{0}}$  &   1.37  &   1.30  &    1.21   & 0.1368                   & \underline{0.0167}     & 0.1232                 \\
$(V-I)_{>r_{0}}$  &   1.31  &   1.26  &    1.17   & 0.3642                   & 0.1937                 & 0.2245                 \\
$C_{<r_{0}}$      &   0.8   &   0.7   &    1.2    & 0.3713                   & \underline{0.0017}     & \underline{0.0032}     \\
$C_{>r_{0}}$      &   0.8   &   1.1   &    1.5    & \underline{0.0296}       & \underline{0.0132}     & 0.1344                 \\
\enddata
\tablecomments{Columns: (1) properties measured in each sample (all values are medians): absolute magnitude in $V$ inside $r_{0}$,
absolute magnitude in $I$ inside $r_{0}$, surface brightness in $V$ inside and outside $r_{0}$,
$V-I$ color inside and outside $r_{0}$, concentration index inside and outside $r_{0}$,
asymmetry level inside and outside $r_{0}$; (2)$-$(4) galaxies in HCG, KPG, and KIG, respectively; (5)$-$(7) 
$P$ values from Dunn's post-tests, where underlined values indicate statistically significant differences.}
\label{tabl10}
\end{deluxetable}

\begin{deluxetable}{lrrrrrr}
\tabletypesize{\scriptsize}
\tablewidth{0pt}
\tablecaption{Median values of Intermediate-type galaxies as measured using $R_e$ $R_e$}
\tablehead{
\colhead{Property}  &  \colhead{HCG}  &  \colhead{KPG}  & \colhead{KIG} &  \colhead{HCG$-$KPG} & \colhead{HCG$-$KIG}  & \colhead{KPG$-$KIG} \\
\colhead{(1)}       & \colhead{(2)}    &  \colhead{(3)}   & \colhead{(4)}  &  \colhead{(5)}     & \colhead{(6)}      & \colhead{(7)}
}
\startdata

$M_V$                & -20.7   & -20.5  & -20.3    & 0.2536  & 0.2375  & 0.5000    \\
$M_I$                & -21.7   & -21.7  & -21.4    & 0.3837  & 0.1465  & 0.2848    \\
$\mu_V$              &  21.1   &  20.9  &  20.5    & 0.1445  & \underline{0.0054}     & 0.0511     \\
$(V-I)$              &  1.23   &  1.17  &  1.09    & 0.3083  & 0.0517  & 0.1417     \\
$R_{90}/R_{50}$      &  2.1    &  2.1   &  2.2     & 0.0714  & 0.1363  & \underline{0.0156}     \\
$R_e$                &  4      &  4     &  3       & 0.2367  & \underline{0.0059}   & 0.0863   \\

\enddata
\tablecomments{ Columns: (1) properties measured in each sample (all values are medians): absolute magnitude in $V$ inside $R_e$,
absolute magnitude in $I$ inside $R_e$, surface brightness in $V$ at $R_e$, $V-I$ color at $R_e$, concentration index, defined by
the ratio of the radii containing 90\% and 50\% of the Petrosian flux, and effective radius in kiloparsec; (2)$-$(4) galaxies in HCG, 
KPG, and KIG, respectively; (5)$-$(7) $P$ values from Dunn's post-tests. Underlined values indicate statistically significant differences.}
\label{tabl11}
\end{deluxetable}

\begin{deluxetable}{lrrrrrr}
\tabletypesize{\scriptsize}
\tablewidth{0pt}
\tablecaption{Properties of Late-type galaxies in different environments}
\tablehead{
\colhead{Property}  &  \colhead{HCGs}  &  \colhead{KPGs}  & \colhead{KIGs} &  \colhead{HCG-KPG} & \colhead{HCG-KIG}  & \colhead{KPG-KIG} \\
\colhead{(1)}       & \colhead{(2)}    &  \colhead{(3)}   & \colhead{(4)}  &  \colhead{(5)}     & \colhead{(6)}      & \colhead{(7)}
}
\startdata

$M_{V}$           &    -19.1  &  -19.2    &   -19.4    & 0.2265               & \underline{0.0290}    & 0.1657     \\
$M_{I}$           &    -20.3  &  -20.7    &   -20.8    & 0.3303               & 0.0560                & 0.1797     \\
$\mu_{<r_{0}}$    &     21.2  &   20.9    &    21.0    & \underline{0.0250}   & 0.1530                & 0.1758     \\
$\mu_{>r_{0}}$    &     22.4  &   22.6    &    22.7    & 0.2172               & 0.2961                & 0.4617     \\
$(V-I)_{<r_{0}}$  &     1.27  &   1.26    &    1.30    & 0.3723               & 0.3455                & 0.1833     \\
$(V-I)_{>r_{0}}$  &     1.27  &   1.15    &    1.30    & 0.4493               & 0.1280                & 0.0794     \\
$C_{<r_{0}}$      &     0.6   &   0.7     &    0.6     & 0.2092               & 0.4323                & 0.1943     \\
$C_{>r_{0}}$      &     0.7   &   0.9     &    0.8     & 0.0963               & \underline{0.0123}    & 0.2601     \\
$A_{<r_{0}}$      &     1.03  &   1.02    &    1.01    & 0.3392               & \underline{0.0007}    & \underline{0.0019}     \\
$A_{>r_{0}}$      &     1.09  &   1.15    &    1.08    & 0.0801               & 0.1400                & \underline{0.0004}     \\

\enddata
\tablecomments{Lines and columns are the same as defined in Table~\ref{tabl7}}
\label{tabl12}
\end{deluxetable}

\begin{deluxetable}{lrrrrrr}
\tabletypesize{\scriptsize}
\tablewidth{0pt}
\tablecaption{Properties of Late-type galaxies as measured using $R_e$}
\tablehead{
\colhead{Property}  &  \colhead{HCG}  &  \colhead{KPG}  & \colhead{KIG} &  \colhead{HCG$-$KPG} & \colhead{HCG$-$KIG}  & \colhead{KPG$-$KIG} \\
\colhead{(1)}       & \colhead{(2)}    &  \colhead{(3)}   & \colhead{(4)}  &  \colhead{(5)}     & \colhead{(6)}      & \colhead{(7)}
}
\startdata

$M_V$                & -20.4   & -19.9  & -20.8    & 0.0707  & 0.0793   & \underline{0.0030}    \\
$M_I$                & -21.7   & -21.0  & -21.9    & 0.0960  & 0.1988   & \underline{0.0035}    \\
$\mu_V$              &  21.4   &  21.5  &  21.6    & 0.4866  & 0.2081   & 0.0890                \\
$(V-I)$              &  1.25   &  1.03  &  1.12    & \underline{0.0156} & 0.1021  & 0.0715     \\
$R_{90}/R_{50}$      &  1.9    &  1.8   &  1.8     & 0.3304  & 0.0996   & 0.1170     \\
$R_e$                &  5      &  3     &  5       & 0.0802  & 0.1448   & \underline{0.0064}  \\

\enddata
\tablecomments{Lines and columns are the same as defined in Table~\ref{tabl12}}
\label{tabl13}
\end{deluxetable}

\clearpage

\clearpage

\clearpage

\onecolumn

\clearpage

\clearpage

\clearpage

\clearpage

\clearpage

\clearpage

\clearpage

\clearpage

\clearpage

\clearpage

\clearpage

\clearpage

\clearpage

\clearpage

\clearpage

\clearpage

\clearpage

\clearpage

\clearpage

\end{document}